%% file: main.tex
\documentclass[12pt,a4paper]{article}

\usepackage{amsmath,amsthm,amssymb,amscd,a4wide}
\usepackage[dvips]{graphicx}

\usepackage{dsfont}
\usepackage{mathrsfs}

\input{def.tex}

\begin{document}

\thispagestyle{empty}
\noindent
ULM-TP/04-3\\
March 2004\\

\vspace*{1cm}

\begin{center}

{\LARGE\bf   Semiclassical propagation of coherent states  \\ 
\vspace*{3mm} with spin-orbit interaction} \\
\vspace*{3cm}
{\large Jens Bolte}%
\footnote{E-mail address: {\tt jens.bolte@physik.uni-ulm.de}}
{\large and Rainer Glaser}%
\footnote{E-mail address: {\tt rainer.glaser@physik.uni-ulm.de}}

\vspace*{1cm}

Abteilung Theoretische Physik\\
Universit\"at Ulm, Albert-Einstein-Allee 11\\
D-89069 Ulm, Germany 
\end{center}

\vfill

\begin{abstract}
We study semiclassical approximations to the time evolution of coherent
states for general spin-orbit coupling problems in two different 
semiclassical scenarios: The limit $\hbar\to 0$ is first taken with fixed 
spin quantum number $s$ and then with $\hbar s$ held constant. In these two 
cases different classical spin-orbit dynamics emerge. We prove that a
coherent state propagated with a suitable classical dynamics approximates
the quantum time evolution up to an error of size $\sqrt{\hbar}$ and
identify an Ehrenfest time scale. Subsequently an improvement of the 
semiclassical error to an arbitray order $\hbar^{N/2}$ is achieved by a 
suitable deformation of the state that is propagated classically.       
\end{abstract}

\newpage

\input{intro.tex}
\input{1sec.tex}   
\input{2sec.tex}

\input{3sec.tex}   
\input{4sec.tex}

\subsection*{Acknowledgments}
A major part of this work has been performed when both authors stayed at 
the Mathe\-matical Sciences Research Institute, Berkeley. We would like 
to thank the MSRI for its hospitality and for the support extended to us. 
Financial support by the Deutsche Forschungsgemeinschaft under contract 
no.\ Ste 241/15-2 is gratefully acknowledged. R.G.\ was also supported through
the Doktorandenstipendium D/02/47460 by Deutscher Akademi\-scher 
Austauschdienst.

\vspace*{0.5cm}
\begin{appendix}
\section*{Appendix: Linear stability of Hamiltonian flows}
\stepcounter{section}
\input{app.stability.tex}
\end{appendix}

{\small
\bibliographystyle{amsalpha}
\bibliography{literatur}}                      

\end{document}

%% file: def.tex
\newcommand{\ud}{\mathrm{d}}
\newcommand{\uD}{\mathrm{D}}
\newcommand{\ui}{\mathrm{i}}
\newcommand{\ue}{\mathrm{e}}

\newcommand{\TRd}{\mathrm{T}^*\rz^d}
\newcommand{\SO}{\mathrm{SO}}
\newcommand{\SU}{\mathrm{SU}}
\newcommand{\SP}{\mathrm{Sp}}
\newcommand{\U}{\mathrm{U}}
\newcommand{\su}{\mathrm{su}}

\newcommand{\Sph}{\mathrm{S}^2}
\newcommand{\Stwo}{\mathrm{S}^2}
\newcommand{\scrS}{\mathscr{S}}

\newcommand{\al}{\alpha}
\newcommand{\be}{\beta}
\newcommand{\de}{\delta}

\newcommand{\eps}{\epsilon}
\newcommand{\ga}{\gamma}

\newcommand{\la}{\lambda}
\newcommand{\om}{\omega}
\newcommand{\Om}{\Omega}
\newcommand{\si}{\sigma}

\newcommand{\vp}{\varphi}
\newcommand{\vr}{\varrho}

\newcommand{\vece}{\boldsymbol e}
\newcommand{\vecn}{\boldsymbol n}
\newcommand{\vecm}{\boldsymbol m}
\newcommand{\vecx}{\boldsymbol x}
\newcommand{\vecs}{\boldsymbol s}
\newcommand{\vecS}{\boldsymbol S}
\newcommand{\vecB}{\boldsymbol B}
\newcommand{\vecC}{\boldsymbol C}
\newcommand{\vecsigma}{\boldsymbol \sigma}

\newcommand{\cD}{{\mathcal D}}
\newcommand{\cH}{{\mathcal H}}

\newcommand{\rz}{{\mathds{R}}}
\newcommand{\nz}{{\mathds{N}}}
\newcommand{\kz}{{\mathds{C}}}

\DeclareMathOperator{\mtr}{tr}

\DeclareMathOperator{\op}{op}

\DeclareMathOperator{\id}{id}

\DeclareMathOperator{\im}{Im}
\DeclareMathOperator{\re}{Re}

\newcommand{\dpr}{\partial}  
\newcommand{\mat}{\mathrm{M}}

\newcommand{\eins}{\mathds{1}}

\newcommand{\nothing}[1]{}

\numberwithin{equation}{section}

\newtheorem{theorem}{Theorem}[section]
\newtheorem{lemma}[theorem]{Lemma}
\newtheorem{prop}[theorem]{Proposition}
\newtheorem{cor}[theorem]{Corollary}

\theoremstyle{definition}


%% file: intro.tex
\section{Introduction}
\label{intro}
Ever since their introduction by Schr\"odinger as early as 1926 \cite{Sch26}, 
coherent states have found an increasing range of applications in quantum
mechanics, see e.g. \cite{KlaSka85,Per86}. In a semiclassical context their 
virtues become particularly transparent in attempts to relate the quantum 
time evolution of a system to its classical trajectories. Coherent states 
can, e.g., even be used to identify the limiting classical dynamics of a 
given quantum system. 

However, apart from the exceptional case of the harmonic oscillator that 
Schr\"odinger chose for his construction, every quantum wave packet 
necessarily disperses. Schr\"o\-dinger's original intention to mimic classical 
trajectories in quantum mechanics can therefore only be put into practice
up to the time scale on which wave packets begin to delocalise. Beyond that 
the quantum time evolution looses its tight relation to classical 
trajectories, although coarser classical structures possibly remain to be 
of influence \cite{SilBee02,Sch04}. 

More recently the notion of an {\it Ehrenfest time} was introduced 
\cite{Chi79,Zas81}, intended to indicate that the Ehrenfest relations can only 
connect quantum dynamics and classical trajectories on limited time scales. 
For classical dynamics with positive Lyapunov exponents it is argued that 
the Ehrenfest time is logarithmic in $\hbar$. This conclusion can be drawn 
from the observation that coherent states are localised in phase space on a 
scale of $\sqrt{\hbar}$, and that an unstable classical dynamics expands 
domains in phase space with exponential rates in the unstable directions. 
Thus for times beyond $\frac{1}{2\la}|\log\hbar|$ a coherent state is no 
longer localised in directions that are expanded with an exponent $\la$. 
A finer analysis reveals that the precise value of the Ehrenfest time 
depends on the problem that is studied; e.g., using $L^2$-norms to measure 
the difference between the quantum time evolution of a coherent state and 
a coherent state that is propagated with the classical dynamics, a critical 
time scale of $\frac{1}{6\la}|\log\hbar|$ was proven to hold \cite{ComRob97}. 
On the other hand, the same difference measured in terms of expectation 
values of observables can be controlled up to times of the order of 
$\frac{1}{2\la}|\log\hbar|$. For details see 
\cite{ComRob97,BonDeB00,BouRob02}. It can moreover be shown that on finite 
time intervals a coherent state is exponentially localised around the 
corresponding classical trajectory \cite{HagJoy00}.

Except for heat kernel asymptotics in the case of particles in non-abelian 
gauge fields \cite{HogPotSch83} most of the previous work on a semiclassical 
control of the propagation of coherent states is concerned with systems that 
possess only translational degrees of freedom. In this article it is our aim 
to extend these investigations to systems with non-relativistic spin-orbit 
interactions. After having identified appropriate coherent states, we intend 
to compare solutions of the  Schr\"odinger equation
\begin{equation*}
 \ui\hbar\frac{\partial\psi}{\partial t}(t,x) = \hat H \psi(t,x) \ ,
\end{equation*}
where the initial wave function $\psi(0)$ is a coherent state, with a coherent 
state that is evolved along suitable classical trajectories. The quantum
Hamiltonians that we wish to allow are of a general spin-orbit coupling type, 
\begin{equation}
\label{eq:soHamilt}
 \hat H = H_0(\hat Q,\hat P) + \vecC(\hat Q,\hat P)\cdot\hat{\vecS} \ ,
\end{equation}
with $\hat Q$, $\hat P$, and $\hat{\vecS}$ denoting the standard position, 
momentum, and spin operators, respectively. Examples of such Hamiltonians
arise when the spin is coupled to an external magnetic field, such that 
$\vecC = \frac{e}{mc}\vecB$, or in the context of atomic spin-orbit coupling 
with $\vecC$ being proportional to orbital angular momentum. 

Apart from atomic and molecular physics spin-orbit coupling also plays an 
important role in nuclei, where it essentially determines their shell structure
\cite{BohMot69}, as well as in solid state physics. In the latter case recent 
experimental progress towards controlling the spin dynamics of electrons 
in semiconductors \cite{DasFabHuZut01} calls for a theoretical description 
of such set-ups. As opposed to some pure quantum calculations semiclassical 
considerations are often particularly transparent and provide a clear physical
picture. With our work we therefore intend to improve the understanding of 
spin-orbit coupling by establishing mathematically rigorous statements about 
the quantum dynamics of localised particles with spin and their relation to 
appropriate classical trajectories. 

One issue to be settled is how the semiclassical limit should be 
performed in the presence of spin-orbit interactions. In principle two 
parameters controlling the passage to a classical description are available, 
which are associated with the two types of degrees of freedom:\ translational
and spin. On the one hand, with (an effective) $\hbar$ approaching zero the 
semiclassical limit is achieved in a standard way for the translational 
degrees of freedom. On the other hand, for an isolated spin $\hbar$ can be 
eliminated from both kinematics and dynamics. The role of a semiclassical 
parameter is then taken over by $1/s$, where $s=1/2,1,3/2,\dots$ denotes the 
spin quantum number. When both types of degrees of freedom interact through 
a spin-orbit coupling one can therefore pass to the semiclassical limit in 
various ways. In the absence of a theory that is uniform in both $\hbar$ and 
$1/s$ we subsequently focus on two important scenarios: 

The most straight forward approach is to view, say, an electron as a particle
with fixed spin $1/2$ and to employ $\hbar$ as the only semiclassical
parameter. In the limit $\hbar\to 0$ the energy scale of the translational 
part $\hat H_0$ in (\ref{eq:soHamilt}) then dominates that of the
spin-orbit coupling term, since in the latter the spin operator $\hat{\vecS}$ 
is proportional to $\hbar$. Although it might appear that thus the spin has 
evaded the leading order semiclassical description, it does in fact contribute
in an essential way through a classical spin precession driven by the orbital 
motion, see \cite{BolKep99a,BolGla00,BolGlaKep01,BolGla02}. E.g., in 
classically chaotic systems this type of spin motion is responsible for the 
quantum eigenvalue spectrum to possess correlations of the Gaussian symplectic 
ensemble of random matrix theory \cite{BolKep99b}. Moreover, in this 
semiclassical framework the exact spectrum of the relativistic hydrogen 
atom is recovered \cite{Kep03a}, and anomalous magneto-oscillations in 
semiconductor devices can be described to a good approximation 
\cite{KepWin02}. 

A second option for the semiclassical limit is to keep the ``classical spin''
$\hbar s$ at a fixed value $S$, thus performing $\hbar\to 0$ and $s\to\infty$
simultaneously. In this scenario the energy scale of spin-orbit interactions
remains comparable to that of the purely translational part, leading to a 
classical spin-orbit Hamiltonian. Therefore, coupled Hamiltonian dynamics
emerge with classical particle trajectories influenced by the spin. This
scenario enables an immediate classical description of the Stern-Gerlach 
experiment, and generally corresponds to a ``strong'' spin-orbit coupling.

In this paper we examine the propagation of coherent states under the
influence of spin-orbit interactions in both of the above mentioned 
semiclassical scenarios. In section~\ref{sec1} we first provide a precise 
characterisation of the quantum Hamiltonians under investigation and then
describe the classical dynamics that will result in due course. 
Section~\ref{sec2} is devoted to outlining the construction of coherent 
states for both translational and spin degrees of freedom, along with their 
basic properties. Our principal results are developed in section~\ref{sec3}. 
For both semiclassical scenarios separately we extend the approach devised 
previously \cite{Hel75,Lit86,ComRob97} in systems without spin in that we 
first construct suitable approximate Hamiltonians that propagate coherent 
states exactly along classical trajectories. We then prove that, measured in 
Hilbert space norm, the full quantum dynamics differs from a classically 
propagated coherent states by an error of size $\sqrt{\hbar}$ as long as 
finite times are taken into account. The vanishing of this difference up to 
some, semiclassically infinite, Ehrenfest time is also established. 
Subsequently we improve the semiclassical error to $O(\hbar^{N/2})$ for 
arbitrary $N\in\nz$ by replacing the classically propagated coherent states 
with a suitable sum of squeezed states. Again such a procedure is possible 
up to the Ehrenfest time. We conclude in section~\ref{sec4} with 
discussing some implications of our main results.


%% file: 1sec.tex
\section{Background}
\label{sec1}
It is our aim to investigate the time evolution of an initial coherent 
state in both translational and spin degrees of freedom generated 
by a general spin-orbit quantum Hamiltonian with an emphasis on a 
semiclassical description. This is the reason why we represent quantum 
observables as matrix valued semiclassical pseudodifferential operators 
within the framework of Weyl calculus, see \cite{Rob87,DimSjo99} for details. 
The quantum Hamiltonians $\hat H$ under consideration are defined on the 
domain $\scrS(\rz^d)\otimes\kz^{2s+1}$ in the Hilbert space 
$L^2(\rz^d)\otimes\kz^{2s+1}$ and are of the form
\begin{equation}
\label{WeylHamilton}
\bigl(\hat H \psi\bigr) (x) = \bigl(\op^W [H]\psi \bigr) (x) =
\frac{1}{(2\pi\hbar)^d}\iint_{\TRd} \ue^{\frac{\ui}{\hbar}\xi\cdot(x-y)}\,
H\Bigl(\frac{x+y}{2},\xi\Bigr)\,\psi(y)\ \ud y\,\ud\xi \ .
\end{equation}
Here $\TRd\cong\rz^d\times\rz^d$ denotes the cotangent bundle over the
euclidean configuration space $\rz^d$, i.e.\ the phase space of the 
translational degrees of freedom. The spin $s=1/2,1,3/2,\dots$ is described 
by the matrix degrees of freedom of the Weyl symbol $H$ and will later be 
represented on its phase space $\Stwo$. Spin-orbit Hamiltonians are 
characterised by symbols of the form
\begin{equation}
\label{eq:WeylSymbol}
H(x,\xi) = H_0(x,\xi) + \hbar\,\vecC(x,\xi)\cdot\ud\pi_s (\vecsigma/2) \ ,
\end{equation}
where $H_0$ and the components $C_k$, $k=1,2,3$, of $\vecC$ are real
valued and smooth functions on $\TRd$ which for all multi-indices $\al$ 
and $\be$ satisfy the growth estimate
\begin{equation}
\label{Symbolclass}
 | \partial_x^\al\partial_\xi^\be F(x,\xi) | \leq
 K_{\al\be}\,\bigl( 1+|x|^2\bigr)^{M_x/2}\bigl(1+|\xi|^2 \bigr)^{M_\xi/2}
\end{equation}
with suitable constants $K_{\al\be}>0$ and $M=(M_x,M_\xi)\in\rz^2$.

The spin-orbit coupling term in (\ref{eq:WeylSymbol}) contains the spin 
operators 
\begin{equation*}
 \hat S_k := \hbar\,\ud\pi_s (\si_k /2) \ , \qquad k=1,2,3 \ , 
\end{equation*}
obeying the well known commutation relations 
$[\hat S_k,\hat S_l]=\ui\hbar\,\eps_{klm}\,\hat S_m$. Here 
\begin{equation}
\label{Paulimat}
 \sigma_1 = \begin{pmatrix} 0&1 \\ 1&0 \end{pmatrix}\ , \quad 
 \sigma_2 = \begin{pmatrix} 0 & - \ui \\ \ui & 0 \end{pmatrix}\ , \quad 
 \sigma_3 = \begin{pmatrix} 1&0\\ 0 &-1 \end{pmatrix} 
\end{equation}
are the Pauli matrices, considered as elements of the real Lie algebra 
$\su(2)$, and $\ud\pi_s$ denotes the $(2s+1)$-dimensional representation of 
$\su(2)$ derived from the corresponding unitary irreducible representation 
$\pi_s$ of the Lie group $\SU(2)$ according to 
$\ud\pi_s(X)=\ui\tfrac{\ud}{\ud\la}\left.\pi_s\bigl(\ue^{-\ui\la X}\bigr)
\right|_{\la=0}$.

The time evolution $\hat U(t)=\ue^{-\frac{\ui}{\hbar}\hat H t}$ generated 
by the quantum Hamiltonian will be unitary provided that $\hat H$ itself is 
essentially self-adjoint on the domain $C_0^\infty (\rz^d)\otimes\kz^{2s+1}$. 
In the present framework this is guaranteed, for sufficiently small $\hbar$, 
once the symbol $H$ is such that $H+\ui$ is elliptic, i.e. if
\begin{equation}
\label{ellipt}
 \bigl\|\bigl(H(x,\xi)+\ui\bigr)^{-1}\bigr\| \leq c\,
 \bigl( 1+|x|^2\bigr)^{-M_x/2}\bigl(1+|\xi|^2 \bigr)^{-M_\xi/2} 
\end{equation}
holds for all $(x,\xi)\in\TRd$ with some constant $c>0$ and $M$ as in 
(\ref{Symbolclass}); here $\|\cdot\|$ is an arbitrary matrix norm. Details 
can be found in \cite{Rob87,DimSjo99}. In the following we assume this 
condition to hold and do not notationally distinguish between $\hat H$ and 
its self-adjoint extension.

In the semiclassical limit we will have to deal with two types of classical
spin-orbit dynamics. In the first case only the translational degrees 
of freedom evolve under a Hamiltonian flow. This is defined on the phase space
$\TRd$ and is generated by the classical Hamiltonian $H_0$. Thus 
$\Phi_0^t(q,p)=\bigl(q(t),p(t)\bigr)$ satisfies Hamilton's equations 
of motion,
\begin{equation*}
 \dot{q}(t) = \partial_\xi H_0\bigl(q(t),p(t)\bigr) \quad\text{and}\quad
 \dot{p}(t) = -\partial_x H_0\bigl(q(t),p(t)\bigr) \ ,
\end{equation*}
with initial conditions $\bigl(q(0),p(0)\bigr)=(q,p)$. This flow then
drives a classical spin through the equations of motion
\begin{equation*}
 \dot{\vecn}(t) = \vecC\bigl(q(t),p(t)\bigr) \times \vecn(t)
\end{equation*}
on the sphere $\Stwo$ with initial condition $\vecn(0)=\vecn$. Here 
$\vecn\in\rz^3$ with $|\vecn|=1$ is considered as a point on $\Stwo$. The 
curve $\vecn(t)$ therefore describes the Thomas precession of a normalised 
classical spin vector on $\Stwo$ along the trajectory 
$\bigl(q(t),p(t)\bigr)$ in $\TRd$. The combined dynamics 
\begin{equation}
\label{eq:skewdyna}
 (q,p,\vecn)\mapsto\bigl(\Phi_0^t(q,p),\vecn(t;q,p,\vecn)\bigr)
\end{equation}
yield a flow on the product phase space $\TRd\times\Stwo$, which is a 
symplectic manifold whose symplectic form is composed of the natural 
symplectic forms of its factors. This flow has the form of a skew product, 
see \cite{CorFomSin82} for details, and thus is not Hamiltonian; however,
it leaves the natural volume measure derived from the symplectic form
invariant.

The second flow relevant for our subsequent discussion includes a classical
spin dynamics coupled to the motion of the translational part in a Hamiltonian
manner and is also defined on the product phase space $\TRd\times\Stwo$. 
These dynamics are generated by the classical spin-orbit Hamiltonian
\begin{equation}
\label{SOHam}
 H_{\mathrm{so}}(x,\xi,\vecn) := H_0 (x,\xi) + S\vecn\cdot\vecC(x,\xi) \ ,
\end{equation}
where the constant $S>0$ measures the length of the classical spin vector 
$\vecs:=S\vecn$. The Hamiltonian flow 
$\Phi^t_{\mathrm{so}}(q,p,\vecn)=\bigl(q(t),p(t),\vecn(t)\bigr)$
is therefore determined by the equations of motion
\begin{equation}
\label{SOflow}
\begin{split}
\dot{q}(t)     &= \partial_\xi 
                  H_{\mathrm{so}}\bigl(q(t),p(t),\vecn(t)\bigr)\ , \\
\dot{p}(t)     &= -\partial_x 
                  H_{\mathrm{so}}\bigl(q(t),p(t),\vecn(t)\bigr) \ , \\
\dot{\vecn}(t) &= \vecC\bigl(q(t),p(t)\bigr) \times \vecn(t)\ .
\end{split}
\end{equation}
The Hamiltonian coupling of the degrees of freedom prescribed by these 
equations imply that in contrast to the previous case the translational 
dynamics are affected by the spin.

Apart from the associated classical flow in semiclassical approximations of
quantum dynamics also the linear stability of the flow plays a role. 
Quantitatively this can be measured in terms of the Lyapunov exponents,
see the appendix for a discussion. They express the rate of phase space
expansion or contraction, respectively, induced by the flow in different 
tangent directions. Moreover, the differential of the flow is a symplectic
map on the tangent bundle of phase space. Its metaplectic representation
is an essential ingredient in the semiclassical propagation of coherent 
states.
 

%% file: 2sec.tex
\section{Coherent states}
\label{sec2}
Within the setting outlined in the preceding section we wish to describe
the time evolution of an initial coherent state semiclassically. The
starting point therefore is the Schr\"odinger equation
\begin{equation*}
 \ui\hbar\,\frac{\partial\psi}{\partial t}(t,x) = \hat H\psi(t,x) \qquad
 \text{with} \qquad \psi(0,x)=\bigl( \vp^B_{(q,p)}\otimes \phi_{\vecn} 
 \bigr)(x)\ , 
\end{equation*}
whose initial condition is the product of a coherent state $\vp^B_{(q,p)}$
of the translational degrees of freedom and a spin-coherent state
$\phi_{\vecn}$. The principal question we then address is to what extent the
quantum mechanical time evolution can be approximated by some classical
dynamics, i.e. we want to estimate the difference
\begin{equation}
\label{Goal}
 \bigl\| \ue^{-\frac{\ui}{\hbar}\hat H t}\bigl(\vp^B_{(q,p)}\otimes 
 \phi_{\vecn}\bigr) - \ue^{\ui \al(t)}\,
 \vp^{B(t)}_{(q(t),p(t))}\otimes \phi_{\vecn(t)} \bigr\| 
\end{equation}
in terms of $\hbar$, where $\bigl(q(t),p(t),\vecn(t)\bigr)$ is an appropriate 
classical trajectory and $\ue^{\ui \al(t)}$ is a suitable phase factor.

For both types of coherent states, $\vp^B_{(q,p)}\in L^2(\rz^d)$ and 
$\phi_{\vecn}\in\kz^{2s+1}$, we use Perelomov's construction \cite{Per86}
that applies to a general Lie group $G$ with unitary irreducible 
representation $\pi$ on a Hilbert space $\cH$:  Fix a non-zero vector 
$\Psi_0\in\cH$ and consider $\Psi_g :=\pi(g)\Psi_0$ for every $g\in G$. 
Hence the vectors $\Psi_g$ and $\Psi_h$ define the same quantum state, i.e. 
$\Psi_h = \ue^{\ui\al}\Psi_g$, if and only if $g^{-1}h$ lies in the stability
subgroup $H\subset G$ of the vector $\Psi_0$,
\begin{equation}
\label{stabH}
 H:=\{ g\in G;\ \pi(g)\Psi_0 = \ue^{\ui\al(g)}\Psi_0 \} \ .
\end{equation}
The quantum states generated by the vectors $\Psi_g$, $g\in G$, can thus be 
labeled by the points $\eta$ of the coset space $G/H$. A section $g(\eta)$ 
in the bundle $G\to G/H$ then determines a choice of vectors 
\begin{equation}
\label{PerelomovCHS}
 \Phi_\eta := \Psi_{g(\eta)} =\pi (g(\eta))\Psi_0 \ ,\qquad\text{for }
 \eta\in G/H\ ,
\end{equation}
representing these states. The vectors $\Phi_\eta$ are called coherent state 
vectors for $(G,\pi,\cH)$.

The two types of coherent states that play a role in the present setting
can be constructed according to this general scheme by choosing the
Heisenberg group $G=H(\rz^d)$ for the translational part and the group
$G=\SU(2)$ for the spin part. We now describe the two situations that emerge 
from this procedure separately.
\subsection{Coherent states for the Heisenberg group}
The Heisenberg group $H(\rz^d)$ is a non-compact $(2d+1)$-dimensional Lie 
group that consists of the elements $(q,p,\la)$ with $(q,p)\in\TRd$ and 
$\la\in\rz$. The group multiplication is given by 
\begin{equation*}
 (q,p,\la) \, (q',p',\la') = \,
 \bigr( q+q',p+p',\la+\la'+\tfrac{1}{2}(pq'-qp') \bigl) \ .
\end{equation*}
According to the Stone-von Neumann Theorem any unitary irreducible 
representation $\pi$ of $H(\rz^d)$ that fulfills 
$\pi(0,0,\la)=\ue^{\frac{\ui}{\hbar}\la}$ is unitarily equivalent to the
Schr\"odinger representation $\rho_\hbar$ on $L^2(\rz^d)$,
\begin{equation*}
 \bigl(\rho_\hbar(q,p,\la)\psi\bigr) (x) = \ue^{\frac{\ui}{\hbar}\la}\, 
 \bigl(\ue^{\frac{\ui}{\hbar}(p\hat Q - q\hat P)}\psi\bigr) (x)
 = \ue^{\frac{\ui}{\hbar}(\la +p(x-\frac{1}{2}q))}\,\psi(x-q) \ .
\end{equation*}
Here $\hat Q_k$ and $\hat P_k$, $k=1,\dots,d$, are the standard self-adjoint 
position and momentum operators defined on suitable domains in $L^2(\rz^d)$. 

In order to construct coherent states for the Heisenberg group we therefore
consider the Schr\"odinger representation $\rho_\hbar$ on $L^2(\rz^d)$. 
One immediately sees that given any non-zero vector $\Psi_0\in L^2(\rz^d)$ 
its stability subgroup is $H=\{(0,0,\la);\ \la\in\rz\}$. Thus coherent 
states can be labeled by points $(q,p)\in G/H$, i.e.\ by points in the 
phase space $\TRd$. This labeling can be achieved in terms of the section 
$g(q,p):=(q,p,-\tfrac{1}{2}qp)$ in $G\to G/H$. One usually prefers a 
reference vector $\Psi_0\in L^2(\rz^d)$ that is normalised, rapidly 
decreasing, and satisfies
\begin{equation*}
\langle \Psi_0,\hat Q\Psi_0 \rangle =0 \qquad\text{and}\qquad
\langle \Psi_0,\hat P\Psi_0 \rangle =0 \ ,
\end{equation*}
so that any reasonable lift of this vector to the phase space $\TRd$ is 
concentrated at $(0,0)\in\TRd$. A convenient choice with these properties is
\begin{equation*}
\psi_0^B (x) := \frac{1}{(\pi\hbar)^{d/4}}\,(\det\im B)^{1/4}\,
\ue^{\frac{\ui}{2\hbar}xBx} \ ,
\end{equation*}
where $B$ is some complex symmetric $d\times d$ matrix with positive-definite
imaginary part. The coherent states (\ref{PerelomovCHS}) that follow from 
the above definitions now read
\begin{equation}
\label{Heisenbcohvec}
\begin{split}
\vp_{(q,p)}^B (x) &= \bigl(\rho_\hbar(q,p,-\tfrac{1}{2}qp)\psi_0^B\bigr)(x)\\
                  &= \frac{1}{(\pi\hbar)^{d/4}}\,(\det\im B)^{1/4}\,
                     \ue^{\frac{\ui}{\hbar}\left(p(x-q)+\frac{1}{2}
                     (x-q)B(x-q)\right)} \ .
\end{split}
\end{equation}
Note that these coherent states differ slightly from more conventional
choices for which $B=\ui\eins_d$ and the section $\tilde g(q,p)=(q,p,0)$ 
are used, leading to a different phase convention. Despite the fact that 
after allowing for more general matrices $B$ the coherent states loose the 
minimum uncertainty property, this generalisation will prove useful since
the action of the metaplectic representation on them can be conveniently
expressed in terms of $B$, see also \cite{Sch01}. The alternative 
phase convention is of less consequence but simplifies the notation.

Although from the above construction it is obvious that a coherent state
$\vp_{(q,p)}^B$ is concentrated in some neighbourhood of the point $(q,p)$ 
in phase space it is instructive to calculate explicit phase-space lifts. 
E.g., its Wigner transform is given by
\begin{equation}
\label{Wignercohvec}
\begin{split}
W[\vp_{(q,p)}^B] (x,\xi) 
   &= \int_{\rz^d}\ue^{-\frac{\ui}{\hbar}\xi y}\,
      \overline{\vp_{(q,p)}^B (x-\tfrac{1}{2}y)}\,
      \vp_{(q,p)}^B(x+\tfrac{1}{2}y) \ \ud y \\
   &= 2^d\,\ue^{-\frac{1}{\hbar}((x,\xi)-(q,p))G_B((x,\xi)-(q,p))}\ , 
\end{split}
\end{equation}
where $G_B$ is the positive-definite symmetric $2d\times 2d$ matrix
\begin{equation*}
 G_B := \begin{pmatrix} \im B+\re B(\im B)^{-1}\re B & -\re B (\im B)^{-1} \\ 
 -(\im B)^{-1}\re B & (\im B)^{-1} \end{pmatrix} \ .
\end{equation*}
This representation reveals a concentration of the coherent state in the
vicinity of the phase-space point $(q,p)$. Moreover, since the sum of position
and momentum uncertainties reads
\begin{equation}
\label{eq:variance}
 \frac{1}{(2\pi\hbar)^d}\iint_{\TRd}\bigl( (x-q)^2 +(\xi-p)^2 \bigr) \,
 W[\vp_{(q,p)}^B] (x,\xi)\ \ud x\,\ud\xi = \frac{\hbar}{2} \bigl( \mtr G_B
 \bigr)^{-1} \ ,
\end{equation}
the spreading of the coherent state in phase space can be measured in terms 
of $G_B$. 
\subsection{Spin-coherent states}
In quantum mechanics the spin of a particle is implemented through the
$(2s+1)$-dimensional irreducible representation $\pi_s$ of the compact Lie 
group $\SU(2)$, where $s=1/2,1,3/2,\dots$ denotes the spin quantum number. 
Within Perelomov's framework spin-coherent states are hence constructed 
from $(\SU(2),\pi_s,\kz^{2s+1})$. The reference vector $\Psi_0\in\kz^{2s+1}$ 
can be chosen such that the coherent states possess the minimum uncertainty 
property; this is achieved with $\Psi_0$ being a maximal weight vector for 
the irreducible representation $\ud\pi_s$ of the Lie algebra $\su(2)$.

The real Lie algebra $\su(2)$ consists of the hermitian and traceless
$2\times 2$ matrices $X$, such that $\ue^{-\ui X}\in\SU(2)$. A convenient
basis of $\su(2)$ is formed by the  Pauli matrices (\ref{Paulimat}). We also 
consider the complexified Lie algebra $\su(2)_\kz:=\su(2)\otimes\kz$ with 
basis given by
\begin{equation*}
 X_\pm:= \frac{1}{2} ( \sigma_1 \pm \ui \sigma_2)\ , \quad 
 X_3:=  \frac{1}{2} \sigma_3 \ ,
\end{equation*}
and commutation relations
\begin{equation*} 
 [X_3,X_\pm]= \pm X_\pm \ , \quad [X_+,X_-]= 2 X_3 \ .
\end{equation*}
The vector $X_3$ spans a Cartan subalgebra, which exponentiates to a 
maximal torus $T\simeq\U(1)$ in $\SU(2)$, and $X_\pm \in \su(2)_\kz$ span 
the root spaces $\mathfrak g_\pm \subset \su(2)_\kz$. Their representations 
$\ud \pi_s(X_\pm)$ are raising and lowering operators, respectively. More 
precisely, the representation space $\kz^{2s+1}$ decomposes into a direct 
sum of the one dimensional eigenspaces of $\ud \pi_s(X_3)$ (weight spaces) 
$V_m = \{\phi \in \kz^{2s+1}; \ \ud \pi_s(X_3)\phi = m\phi \}$, where 
$m=-s,-s+1,\dots,s$. The raising and lowering operators $\ud \pi_s(X_\pm)$ 
map the weight spaces into one another, $\ud\pi_s (X_\pm) V_m = V_{m\pm1}$
for $m\neq\pm s$. The weights $m=\pm s$ are called maximal and minimal 
weights, respectively. The corresponding weight vectors are annihilated by 
the raising or lowering operator. In the usual angular momentum notation 
a normalised weight vector is denoted as $|s,m \rangle$. 

For a given representation $\pi_s$ of $\SU(2)$ we choose a maximal weight
vector $|s,s \rangle$ as the reference vector $\Psi_0$. According to 
(\ref{stabH}) the stability group of this vector is
\begin{equation*}
 H = \{ g=\ue^{-\ui \la\si_3};\ \la\in [0,2\pi) \} \cong \U(1) \ ,
\end{equation*}
which can be identified with a maximal torus $T$. Thus coherent states are
labeled by points in the coset space 
\begin{equation*}
 G/H \cong \SU(2)/\U(1) \cong \Sph \  .
\end{equation*}
As in the case of the Heisenberg group this manifold is naturally 
symplectic and can be viewed as the corresponding classical phase space. 

The definition of coherent states finally requires a section in $G\to G/H$, 
i.e.\ in the Hopf bundle $\SU(2)\to\Sph$. This principal $\U(1)$-bundle, 
however, is non-trivial so that no smooth global section exists. We therefore 
here give local constructions that, nevertheless, allow for suitable 
interpretations in terms of global objects. We parameterise points on 
$\Sph$ by $\vecn\in\rz^3$ with $|\vecn|=1$ and use spherical coordinates,
$\vecn(\theta,\vp)=(\sin\theta\cos\vp,\sin\theta\sin\vp,\cos\theta)$ with
$\theta\in[0,\pi)$ and $\vp\in[0,2\pi)$. Introducing 
$\vece_\vp := (-\sin\vp,\cos\vp,0)$ our choice of a local section reads 
(see also \cite{Per86})
\begin{equation*}
 g_{\vecn} = \ue^{-\frac{\ui}{2}\theta\vece_\vp\cdot\vecsigma} =
 \begin{pmatrix} 
 \cos\tfrac{\theta}{2} & -\sin\tfrac{\theta}{2}\,\ue^{-\ui\vp} \\
 \sin\tfrac{\theta}{2}\,\ue^{\ui\vp} & \cos\tfrac{\theta}{2}
 \end{pmatrix} \ .
\end{equation*}
Under the double covering map $R:\SU(2)\to\SO(3)$ that is defined through 
$(R(g)\vecx)\cdot\vecsigma = g\vecx\cdot\vecsigma g^{-1}$, the matrix 
$g_{\vecn(\theta,\vp)}$ corresponds to the rotation $R(g_{\vecn(\theta,\vp)})$
about the axis $\vece_\vp$ with angle $-\theta$, such that 
$R(g_{\vecn})\vece_3 = \vecn$, where $\vece_3 =(0,0,1)$ represents the north
pole on $\Stwo$. With these choices spin-coherent states are the normalised 
vectors
\begin{equation}
\label{SU2chsdef}
\phi_{\vecn} := \pi_s(g_{\vecn})|s,s\rangle \ .
\end{equation}
These states are conveniently represented on the phase space $\Sph$ through
the Husimi transform (see \cite{Per86}),
\begin{equation*}
 h[\phi_{\vecn}](\vecm) := |\langle \phi_{\vecm},\phi_{\vecn} \rangle| =
 \left( \frac{1+\vecm\cdot\vecn}{2} \right)^s \ ,
\end{equation*}
which clearly indicates a concentration, in the semiclassical limit 
$s\to\infty$, of $\phi_{\vecn}$ at the point $\vecn\in\Sph$. 

Our next aim is to investigate the relation between the propagation of
a spin-coherent state (\ref{SU2chsdef}) generated by a (time-dependent, 
linear) spin-Hamiltonian
\begin{equation}
\label{spinHam}
\hat H_{\text{spin}} = \vecC(t)\cdot\hat{\vecS} 
\end{equation}
defined on $\kz^{2s+1}$, and a suitable classical time evolution $\vecn(t)$
on $\Sph$. Here $\hat{\vecS}$ denotes the vector of spin operators 
$\hat S_k = \hbar\,\ud\pi_s(\si_k/2)$. The dynamics of a coherent state 
$\phi_{\vecn}$ follows from the equation
\begin{equation}
\label{spineq}
\ui\hbar\frac{\partial\phi}{\partial t}(t) = \hat H_{\text{spin}}\phi(t) 
\qquad\text{with}\qquad \phi(0)=\phi_{\vecn} \ .
\end{equation}
A solution of this problem can be related to the curve $g(t)$, $t\in\rz$,
in $\SU(2)$ determined by
\begin{equation}
\label{groupeq}
\dot{g}(t) + \tfrac{\ui}{2}\,\vecC(t)\cdot\vecsigma\,g(t) = 0 
\qquad \text{with}\qquad g(0)=\id_{\SU(2)}
\end{equation}
through 
\begin{equation}
\label{spinprop}
\phi(t) = \pi_s(g(t))\phi(0) = \pi_s\bigl(g(t)g_{\vecn}\bigr) |s,s\rangle \ . 
\end{equation}
An associated classical time evolution then arises from the adjoint action 
of $g(t)$ on $\vecn\cdot\vecsigma\in\su(2)$ via 
$\vecn(t)\cdot\vecsigma := g(t)\vecn\cdot\vecsigma g(t)^{-1}=(R(g(t))\vecn)
\cdot\vecsigma$. This implies
\begin{equation}
\label{preceq}
\dot{\vecn}(t) = \vecC(t)\times\vecn(t) \qquad\text{with}\qquad
\vecn(0)=\vecn \ .
\end{equation}
The corresponding coherent state vector $\phi_{\vecn(t)}$ differs from the 
quantum time evolution $\phi(t)$ of $\phi_{\vecn}$ only by a phase; both
vectors therefore describe the same quantum state. 

Since this phase is required for later purposes, we now determine it 
explicitly. To this end we notice that $\vecn(t)$ can on the one hand be
represented as
\begin{equation*}
 \vecn(t) =  R\bigl(g(t)\bigr)\vecn = R\bigl(g(t)g_{\vecn}\bigr)\vece_3 \ ,
\end{equation*}
and on the other hand as
\begin{equation*}
 \vecn(t) = R\bigl(g_{\vecn(t)}\bigr)\vece_3 \ .
\end{equation*}
Thus, under the double covering map, $g_{\vecn(t)}^{-1}g(t)g_{\vecn}\in\SU(2)$ 
is associated with a rotation about $\vece_3$ with some angle $\vr(t)$,
such that
\begin{equation*}
 g_{\vecn(t)}^{-1}g(t)g_{\vecn} = \ue^{\frac{\ui}{2}\vr(t)\si_3} \in T\ .
\end{equation*}
From (\ref{spinprop}) it now follows that 
\begin{equation}
\label{spinprop1}
\phi(t) = \pi_s\bigl(g(t)g_{\vecn}\bigr) |s,s\rangle 
        = \pi_s\bigl(g_{\vecn(t)}\bigr)
           \pi_s\bigl(\ue^{\frac{\ui}{2}\vr(t)\si_3}\bigr) |s,s\rangle 
        = \ue^{\ui s\vr(t)}\,\phi_{\vecn(t)} \ ,
\end{equation}
thus confirming the claimed relation between the quantum and `classical'
propagation of the spin-coherent state $\phi_{\vecn}$. Due to the explicit
dependence of the phase on $s$ it suffices to calculate the angle $\vr(t)$
for $s=\tfrac{1}{2}$. For this one notices that in polar coordinates
\begin{equation}
\label{chs1/2}
\phi(t) = \ue^{\frac{\ui}{2}\vr(t)}\,
          g_{\vecn(t)}|\tfrac{1}{2},\tfrac{1}{2}\rangle 
        = \begin{pmatrix} 
          \cos(\tfrac{\theta(t)}{2})\,\ue^{\frac{\ui}{2}\vr(t)} \\ 
          \sin(\tfrac{\theta(t)}{2})\,\ue^{\ui(\vp(t)+\frac{1}{2}\vr(t))}
          \end{pmatrix} \ .
\end{equation}
In a standard calculation (see e.g.\ \cite{BolKep99b}) $\vr(t)$ 
can now be determined 
by using (\ref{chs1/2}) in equation (\ref{spineq}), leading to  
\begin{equation}
\label{deltaoft}
\vr(t) = - \int_0^t \Bigl( \vecC(t')\cdot\vecn(t') + 
\big(1-\cos\theta(t')\bigr)\dot{\vp}(t')\Bigr)\ \ud t' \ .
\end{equation}
If one introduces a classical spin vector $\vecs := S\vecn$, with some
$S>0$, one can relate the angle $\vr(t)$ to Hamilton's principal 
function of the spin. The observation that 
\begin{equation*}
 L_{\text{spin}}(t) = -\vecC(t)\cdot\vecs(t) - S\,\big(1-\cos\theta(t)\bigr)
 \dot{\vp}(t)
\end{equation*}
is the Lagrangean of the classical spin motion implies $S\vr(t)$ to be the
spin-action $R_{\mathrm{spin}}(t)$.

%% file: 3sec.tex
\section{Time evolution of coherent states}
\label{sec3}
In this section we discuss the time evolution of coherent states in 
two different semiclassical limits. In the first scenario we consider
$\hbar\to 0$ while the spin quantum number $s$ is fixed. This will imply
that primarily the translational degrees of freedom become semiclassical. 
The spin-orbit interaction therefore occurs on the level of the subprincipal 
symbol of the Hamiltonian (\ref{eq:WeylSymbol}), enforcing the skew-product 
structure (\ref{eq:skewdyna}) of the resulting classical dynamics with the 
translational motion driving the spin. 

In the second scenario we fix the product $S:=\hbar s$ and hence consider 
the combined limits $\hbar\to 0$ and $s\to\infty$. Thus both types of
degrees of freedom are treated semiclassically on equal footing. This
results in a classical spin-orbit coupling with the Hamiltonian dynamics
(\ref{SOflow}) generated by the function (\ref{SOHam}).

We begin with the first scenario which is close to the time evolution of 
coherent states without spin degrees of freedom.
\subsection{Semiclassics with fixed spin}
In the present scenario $\hbar$ is the only semiclassical parameter so
that we consider the quantum Hamiltonian (\ref{WeylHamilton}) as a Weyl 
operator with matrix valued symbol (\ref{eq:WeylSymbol}) that has a scalar 
principal part; the subprincipal symbol then contains the spin-orbit
coupling. This setting ensures that the propagation of coherent states
is closely analogous to the case without spin, compare \cite{ComRob97}.

Guided by this analogy we first construct an approximate Hamiltonian that 
propagates coherent states exactly. Regarding the translational part we 
exploit the fact that the time evolution generated by a quadratic Hamiltonian 
preserves the form $\vp^B_{(q,p)}$ given in (\ref{Heisenbcohvec}) of a 
coherent state for the Heisenberg group. The spin part of the coherent 
state shall be propagated by a Hamiltonian of the form (\ref{spinHam}) 
and can hence be calculated explicitly. Using the convenient notation 
$w:=(x,\xi)\in\TRd$, we now consider the Taylor expansion of the symbol 
(\ref{eq:WeylSymbol}) about some smooth curve $z(t)=\bigl(q(t),p(t)\bigr)$ 
in phase space. The Weyl quantisation of the leading terms in the Taylor 
expansion (of different order in the principal and in the subprincipal 
symbol),
\begin{equation} 
\label{eq:QuadrExpansH}
 H_Q(t,w):= \sum_{|\nu|=0}^2 \frac{1}{\nu !} H_0^{(\nu)}\bigl(z(t)\bigr) 
            \bigl(w-z(t)\bigr)^\nu + \hbar\,\vecC\bigr(z(t)\bigl) \cdot 
            \ud\pi_s (\vecsigma/2) \ ,
\end{equation}
yields a quantum Hamiltonian $\hat H_Q(t)$ that is quadratic in $\hat Q$ 
and $\hat P$ and linear in $\hat{\vecS}$. Here $H_0^{(\nu)}(w)$ stands for the 
derivative $\partial^\nu_w H_0(w)$ of order $|\nu|$ in the $2d$ components of 
$w=(x,\xi)$. The time evolution $\psi_Q(t)\in L^2(\rz^d)\otimes\kz^{2s+1}$ 
of a coherent state $\vp^B_{(q,p)}\otimes\phi_{\vecn}$ generated by the 
approximate Hamiltonian,
\begin{equation} 
\label{eq:CauchyProblH2}
 \ui\hbar\frac{\partial\psi_Q}{\partial t}(t) = \hat H_Q (t)\psi_Q (t)
 \qquad\text{with}\qquad \psi_Q (0) = \varphi_{(q,p)}^B \otimes \phi_{\vecn}\ ,
\end{equation}
can be expressed in terms of a coherent state:
\begin{prop} 
\label{prop:ExactProp}
The solution of the quadratic Schr\"odinger equation (\ref{eq:CauchyProblH2}) 
is a time-dependent coherent state with an additional phase,
\begin{equation}  
\label{eq:SolnCauchyProblH2}
 \psi_Q (t) = \ue^{\ui\left( \frac{R_0(t)}{\hbar} + s\vr(t) 
              + \frac{\pi}{2}\sigma(t)\right)}\,\vp^{B(t)}_{(q(t),p(t))} 
              \otimes \phi_{\vecn(t)} \ . 
\end{equation} 
Here $\bigl(q(t),p(t)\bigr)=\Phi_0^t(q,p)$ is the solution of Hamilton's 
equations of motion generated by the principal symbol $H_0$,
\begin{equation}
\label{Heq}
 \dot q(t) = \dpr_{\xi} H_0\bigl(q(t),p(t)\bigr)\ , \qquad \dot p(t) = 
 - \dpr_x H_0\bigl(q(t),p(t)\bigr)\ , 
\end{equation}
with initial condition $\bigl(q(0),p(0)\bigr)=(q,p)$ and principal function
\begin{equation}
\label{principalfct}
 R_0(t) = \int_0^t\bigl( p(t')\dot q(t') - H_0(q(t'),p(t'))\bigr) \ \ud t'\ .
\end{equation}
The complex symmetric $d\times d$ matrix B(t) is given by 
\begin{equation}
\label{Bsolve}
 B(t) = \bigl( \dpr_q q(t) B + \dpr_p q(t) \bigr)
        \bigl( \dpr_q p(t) B + \dpr_p p(t) \bigr)^{-1} \ ,
\end{equation}
where the derivatives are taken with respect to the initial conditions; it 
also gives rise to the Maslov phase $\sigma(t)$. Moreover, $\vecn(t)$ is a 
solution of the spin precession equation (\ref{preceq}) in which $\vecC(t)$ 
stands for $\vecC\bigl(q(t),p(t)\bigr)$ from (\ref{eq:QuadrExpansH}); $\vr(t)$ 
then is the associated angle (\ref{deltaoft}).
\end{prop}
\begin{proof}
For the proof we adapt the method of \cite{Sch01} to the present situation 
and therefore introduce the ansatz 
\begin{equation*}
 \psi_Q(t,x) = (\pi \hbar)^{-d/4}\,\ga(t)\,\ue^{\frac{\ui}{\hbar} \theta(t)} 
 \,\ue^{\frac{\ui}{\hbar}\left( p(t)(x-q(t))+\frac{1}{2}(x-q(t))B(t)(x-q(t)) 
 \right)}\,\phi_{\vecn(t)}
\end{equation*}
in equation (\ref{eq:CauchyProblH2}). To deal with the spin contribution 
to the left-hand side we use the fact that according to (\ref{spineq}) and 
(\ref{spinprop1}) 
\begin{equation}
\label{spinderivative}
 \ui\hbar\frac{\partial}{\partial t}\bigl(\ue^{\ui s\vr(t)}\,\phi_{\vecn(t)} 
 \bigr) = \hbar\,\vecC\bigl(q(t),p(t)\bigr)\cdot\ud\pi_s(\vecsigma/2)\,
 \ue^{\ui s\vr(t)}\,
 \phi_{\vecn(t)}\ ,
\end{equation}
if and only if $\vecn(t)$ solves (\ref{preceq}). It hence remains to consider 
(see \cite{Sch01})
\begin{equation}
\label{transeval}
\begin{split}
 \ui\hbar\frac{\partial}{\partial t}&\left( (\pi \hbar)^{-d/4}\,\ga(t)\,
 \ue^{\ui \left(\frac{\theta(t)}{\hbar}-s\vr(t)\right)}\,
 \ue^{\frac{\ui}{\hbar}\left( p(t)(x-q(t))+\frac{1}{2}(x-q(t))B(t)(x-q(t))
 \right)}\right)\ue^{\ui s\vr(t)}\,\phi_{\vecn(t)} \\
   &= \left[ H_0 + H'_{0,x}(x-q(t))+H'_{0,\xi}\cdot B(t)(x-q(t))+\frac{1}{2}
      (x-q(t))\cdot H''_{0,xx}(x-q(t))\right. \\
   &\qquad +\frac{1}{2}(x-q(t))\cdot H''_{0,\xi x}B(t)(x-q(t)) +
      \frac{1}{2}(x-q(t))\cdot B(t)H''_{0,\xi x}(x-q(t)) \\
   &\qquad \left. +\frac{1}{2}(x-q(t))\cdot B(t)H''_{0,\xi\xi}B(t)(x-q(t)) +
       \frac{\hbar}{2\ui}\mtr\bigl( H''_{0,\xi x} + H''_{0,\xi\xi}B(t) \bigr) 
       \right]\,\psi_Q(t,x) \  .
\end{split}
\end{equation}
Here the abbreviations 
$H'_{0,x}=\partial_x H_0$ and $H''_{0,\xi x}=\partial_\xi\partial_x H_0$, 
etc.\  have been employed. These expressions are to be evaluated at $z(t)$. 
Comparing coefficients of powers of $\hbar$ and of $\bigl(x-q(t)\bigr)$ in 
(\ref{transeval}) then yields the conditions
\begin{equation} 
\begin{split}
\label{conditions}
 \dot{\theta} 
    &= \dot{q}p - H_0 \ , \\
 - \dot{p} + B \dot{q} 
    &= H'_{0,x} + BH'_{0,\xi} \ , \\
 - \dot{B}
    &= H''_{0,xx} + H''_{0,\xi x}B + BH''_{0,\xi x} + BH''_{0,\xi\xi}B \ , \\
 \frac{\dot{\ga}}{\ga} 
    &=  -\frac{1}{2} \mtr\bigl( H''_{0,\xi x} + H''_{0,\xi\xi} 
        B\bigr) + \ui s\dot{\vr}  \ .
\end{split} 
\end{equation} 
With the identification $R_0=\theta$ the first and the second equation  
immediately imply (\ref{principalfct}) and (\ref{Heq}), respectively. 

The other two equations involve the time evolution $B(t)$ of the complex 
symmetric $d\times d$ matrix $B$ with positive-definite imaginary part; they 
determine the action of the metaplectic group on the vector $\vp^B_{(q,p)}$.
At this stage we recall that the symplectic group $\SP(d,\rz)$ acts on the 
Siegel upper half-space (see \cite{Fol89})
\begin{equation*}
 \Sigma_d := \{ Z\in\mat_d (\kz);\ Z^{\mathrm{T}} =Z,\ \im Z >0 \} 
\end{equation*}
via
\begin{equation*}
 S[Z] = (S_{11}Z+S_{12})(S_{21}Z+S_{22})^{-1} \ , \quad\text{where}\quad
 S = \begin{pmatrix} S_{11} & S_{12} \\  S_{21} & S_{22} \end{pmatrix}
 \in\SP(d,\rz) \ .
\end{equation*}
In the present context the differential of the Hamiltonian flow $\Phi_0^t$
generated by the classical Hamiltonian $H_0$ is symplectic, 
$S_{0,z}(t):=\uD\Phi_0^t(z)\in\SP(d,\rz)$, and hence can act on the initial 
value $B\in\Sigma_d$. Indeed,
\begin{equation}
\label{SofB}
 B(t) = S_{0,z}(t)[B] \in \Sigma_d
\end{equation}
yields the solution of the third equation in (\ref{conditions}) and implies
(\ref{Bsolve}). The fourth equation requires the introduction of the Maslov 
multiplier $m(S,Z):=\bigl(\det(S_{21}Z+S_{22})\bigr)^{-1/2}$ for 
$S\in\SP(d,\rz)$ and $Z\in\Sigma_d$. This allows us to define the Maslov phase 
$\si(t)$ through $\ue^{\ui\frac{\pi}{2}\si(t)}=m(O(t),i\eins)$, where $O(t)$ 
is an orthogonal symplectic matrix that is uniquely associated with 
$S_{0,z}(t)$. One can then show (cf. \cite{Sch01}) that
\begin{equation*}
 \gamma(t) = \bigl( \det\im B(t) \bigr)^{1/4} \, 
             \ue^{\ui\frac{\pi}{2}\si(t) + \ui s\vr(t)} \ .
\end{equation*}
\end{proof}
We remark that the state (\ref{eq:SolnCauchyProblH2}) is closely analogous
to the respective solution without spin-orbit coupling. It differs from
the latter only by the factor $\ue^{\ui s\vr(t)}\phi_{\vecn(t)}$. This
observation not only means that quantum mechanically the translational
part and the spin part are not entangled, but also on the classical level 
the translational dynamics are independent of the spin precession $\vecn(t)$. 
The combination of classical translational and spin motion rather has the 
structure of a skew product (\ref{eq:skewdyna}), indicating that only the 
spin dynamics depends on the translational part, and not vice versa.

Our aim now is to compare the time evolution generated by the original 
quantum Hamiltonian $\hat H$ with the one generated by the approximate
Hamiltonian $\hat H_Q(t)$. For this we will follow the method devised in
\cite{ComRob97} for the case without spin. The presence of spin 
requires some modifications that, however, are modest when the spin 
quantum number $s$ is fixed. But for the clarity of the presentation, and 
to prepare for the more involved situation to be dealt with in the second
semiclassical scenario, we will now present the argument in some detail. 

As stated in section~\ref{sec1} the Hamiltonian $\hat H$ generates a unitary
and strongly continuous one-parameter group $\hat U(t,t_0)$, if its symbol 
satisfies the ellipticity condition (\ref{ellipt}). When considering the 
limit $\hbar\to 0$ and keeping $s$ fixed this requirement need only be
imposed on the principal symbol, i.e.\ we demand
\begin{equation}
\label{ellipt1}
 |H_0(x,\xi)+\ui| \geq c\,
 \bigl( 1+|x|^2\bigr)^{M_x/2}\bigl(1+|\xi|^2 \bigr)^{M_\xi/2} \ . 
\end{equation}
Let now $\hat U_Q(t,t_0)$ be the corresponding unitary group generated by 
$\hat H_Q(t)$. Using Duhamel's principle we may then express the difference 
between these unitary operators as
\begin{equation}
\label{Duhamel}
 \hat U(t,t_0) - \hat U_Q(t,t_0) = \frac{1}{\ui\hbar}\int_{t_0}^t
 \hat U(t,t') \bigl(\hat H-\hat H_Q(t')\bigr) \,\hat U_Q(t',t_0)\ \ud t' \ .
\end{equation}
Since we are interested in the difference (\ref{Goal}), we have to consider
the action of (\ref{Duhamel}) on the initial state 
$\vp^B_{(q,p)}\otimes\phi_{\vecn}$ with $t_0 =0$. This requires an estimate of
\begin{equation} 
\label{eq:DiffHamiltonians}
 \| (\hat H - \hat H_Q(t')) \psi_Q(t') \| \ , 
\end{equation}
where $\psi_Q(t')$ is the time dependent coherent state 
(\ref{eq:SolnCauchyProblH2}). One can achieve this with the help of the 
following lemma, which is an immediate extension of a result given in 
\cite{ComRob97}.
\begin{lemma} 
\label{lem:ExpdedWeylOpOnS}
Let $f,g \in C^\infty(\TRd)$ be symbols that satisfy the estimate 
(\ref{Symbolclass}) with $M=0$ and let $F:\TRd\to\TRd$ be a linear map with
Hilbert-Schmidt norm $\|F\|_{\mathrm{HS}}$. Fix $\al,\be\in\nz^{2d}$ with 
$k:=|\al|=|\be|+2>2$ and introduce the symbol
\begin{equation*}
 A(w):= (Fw)^\al f(Fw) + \hbar\, (Fw)^\be g(Fw) \ . 
\end{equation*}
Then for any real number $\kappa >0$ there exist $C>0$ 
and $N\in\nz$ such that 
\begin{equation*} 
 \| \op^W[A] \psi_{\hbar} \| \le C \hbar^{k/2} \left( \|F\|_{\mathrm{HS}}^k 
 \sup_{|\gamma| \le k +N} |\dpr_w^\gamma f(w)| + \|F\|_{\mathrm{HS}}^{k-2} 
 \sup_{|\gamma| \le k-2+N} |\dpr_w^\gamma g(w)| \right)
\end{equation*}
holds for any function 
$\psi_\hbar(x)=\hbar^{-d/4}\psi\bigl(x/\sqrt{\hbar}\bigr)$ with 
$\psi\in\scrS(\rz^d)$ and $0<\hbar +\sqrt{\hbar}\|F\|_{\mathrm{HS}}<\kappa$.
\end{lemma}
We intend to apply this lemma to the difference (\ref{eq:DiffHamiltonians}), 
with $f$ corresponding to the Taylor remainder of $H_0$ of order three and 
$g$ to the Taylor remainder of $\vecC\cdot\ud\pi_s(\vecsigma/2)$ of order one.
But first we replace (\ref{eq:DiffHamiltonians}) by
\begin{equation}
\label{eq:differenceconj}
 \| \hat U_Q(t',0)^\ast \bigl(\hat H - \hat H_Q(t')\bigr) 
 \hat U_Q(t',0) \psi_Q(0) \| 
\end{equation}
and invoke an appropriate Egorov theorem. Since the Hamiltonian generating
$\hat U_Q(t,0)$ has a symbol that is composed of a scalar and quadratic
principal part as well as a matrix valued subprincipal part, one can 
combine the techniques used in \cite{BolGla00} and \cite{Sch01}. This 
shows that 
\begin{equation}
\label{Wdef}
 \hat W(t) := \hat U_Q(0,t) \bigl( \hat H - \hat H_Q(t)\bigr) \hat U_Q(t,0)
\end{equation}
is a Weyl operator with symbol
\begin{equation}
\label{eq:QEgorov}
 W(t,w) = d^\ast\bigl(z(t)\bigr) \, \bigl( H-H_Q(t)\bigr)
 \bigl(z-S_{0,z}^{-1}(t)(w-z(t))\bigr) \, d\bigl(z(t)\bigr) \ .
\end{equation}
Here $d\bigl(z(t)\bigr)$ is the representation $\pi_s\bigl(g(t)\bigr)$ of 
the solution to equation (\ref{groupeq}) in which $\vecC(t)$ stands for 
$\vecC\bigl(z(t)\bigr)$. Thus 
\begin{equation*}
 d\bigl(z(t)\bigl)\phi_{\vecn} = \ue^{\ui s\vr(t)}\,\phi_{\vecn(t)} 
\end{equation*}
describes the transport of a spin-coherent state along the trajectory
$z(t)$. Since the principal part of the symbol $H-H_Q(t)$ is scalar it
is not affected by the conjugation with $d\bigl(z(t)\bigl)$. In the
subprincipal term this conjugation rotates the spin operator 
$\hat{\vecS}=\hbar\ud\pi_s(\vecsigma/2)$ to $R\bigl(g(t)\bigr)\hat{\vecS}$.
Therefore, the spin part of the Egorov relation (\ref{eq:QEgorov}) does 
not contribute to an estimate of (\ref{eq:differenceconj}) in an essential
way.  

If one now localises the symbol (\ref{eq:QEgorov}) in $w$ with some
smooth function that is compactly supported around $z(t)$, leading to an
error of size $O(\hbar^\infty)$ when one applies $\hat W$ to a coherent 
state located at $z(t)$, one can proceed to use 
Lemma~\ref{lem:ExpdedWeylOpOnS} as in 
\cite{ComRob97}. This shows that there exists a constant $K >0$ such that
\begin{equation}
\label{firstest}
 \| (\hat H - \hat H_Q(t)) \psi_Q(t) \| \le K \hbar^{3/2}
 \theta(t)^3 \, \delta(t)^m  \  , 
\end{equation}
where
\begin{equation}
\label{thetdeldef}
 \theta(t):= \max\Bigl\{ 1, \sup_{t'\in [0,t]} \| S_{0,z}(t') \|_{\mathrm{HS}} 
 \Bigr\} \quad\text{and}\quad  
 \delta(t):= \sup_{t'\in [0,t]} \bigl( 1+|z(t')| \bigr) 
\end{equation}
depend on the classical trajectory $z(t)=(q(t),p(t))$. The constant 
$m=\max\{M_x,M_\xi\}$ is related to $M=(M_x,M_\xi)$ appearing in 
(\ref{Symbolclass}). We then obtain: 
\begin{theorem}
\label{prop:1stapprox}
Let the conditions imposed on the Hamiltonian in section~\ref{sec1} and
the ellipticity condition (\ref{ellipt1}) hold. Then the coherent state 
$\psi_Q(t)$ defined in (\ref{eq:SolnCauchyProblH2}) semiclassically 
approximates 
$\psi(t)=\hat U(t,0) \bigl( \vp_{(q,p)}^B \otimes\phi_{\vecn}\bigr)$ 
in the following sense,
\begin{equation}
\label{quadapprox}
 \| \psi(t) - \psi_Q(t) \| \le K\sqrt{\hbar}\,t\,\theta(t)^3 \  .
\end{equation}
The right-hand side vanishes in the combined limits $\hbar\to 0$ and 
$t\to\infty$ as long as $t\ll T_z(\hbar)$. The time scale $T_z(\hbar)$ 
depends on the linear stability of the trajectory $z(t)$. If the latter 
possesses a positive and finite maximal Lyapunov exponent 
$\la_{{\mathrm{max}}}(z)$, one has 
$T_z(\hbar)=\frac{1}{6\la_{{\mathrm{max}}}(z)}|\log\hbar|$. In the case 
of a trajectory on a (non-degenerate) KAM-torus this time scale is 
$T_z(\hbar)=C\,\hbar^{-1/8}$.
\end{theorem}
\begin{proof}
Conservation of energy, $H_0\bigl(z(t)\bigr) = E$, together with the
ellipticity condition (\ref{ellipt1}) implies that $\de(t)$ is bounded from
above by some constant depending on $E$. Thus the estimate (\ref{firstest}) 
immediately yields (\ref{quadapprox}) when used in (\ref{Duhamel}). 

If $z(t)$ is a trajectory with a positive, but finite, maximal Lyapunov
exponent the dominant behaviour as $t\to\infty$ comes from the term 
$\theta(t)^3$. This is due to the relation
\begin{equation*}
 \la_{{\mathrm{max}}} (z) = \limsup_{t\to\infty} \frac{1}{t}
 \log\|S_{0,z}(t)\|_{\mathrm{HS}} \ ,
\end{equation*}
see (\ref{MaxLyapunov}), which readily implies 
$T_z(\hbar)=\frac{1}{6\la_{{\mathrm{max}}}(z)}|\log\hbar|$. In the
appendix we also discuss sufficient conditions under which finite maximal 
Lyapunov exponents occur.

If $z(t)$ is a trajectory on a KAM-torus one can introduce local action-angle
variables $(I,\phi)$ in a neighbourhood of that torus such that in these
canonical coordinates the flow reads $I(t)=I$ and $\phi(t)=\phi+\omega(I)t$,
see \cite{Laz93}. One therefore finds
\begin{equation*}
 \|S_{0,z}(t)\|^2_{\mathrm{HS}} = 2d + f(I)\,t^2 \ ,
\end{equation*}
such that $\theta(t)\sim Kt$ as $t\to\infty$, which finally yields 
$T_z(\hbar)=C\,\hbar^{-1/8}$. In the degenerate case, where $f(I)=0$,
this changes to $T_z(\hbar)=C\,\hbar^{-1/2}$.
\end{proof}
In a next step we want to improve the semiclassical error in 
(\ref{quadapprox}) to an arbitrary (half-integer) power of $\hbar$. This 
requires higher order approximations that may be achieved as in 
\cite{ComRob97} by iterating Duhamel's principle (\ref{Duhamel}), resulting 
in the Dyson expansion
\begin{equation}
\begin{split} 
\label{eq:Dyson}
 \hat U(t,0)  - \hat U_Q(t,0) 
 &= \sum_{j=1}^{N-1} (\ui \hbar)^{-j} \int_{0}^t \dots \int_{t_{j-1}}^t 
     \hat U_Q(t,0)\,\hat W(t_j)\dots\hat W(t_1)\   
    \ud t_j \dots \ud t_{1} \\
 &\qquad  +R_N(t;\hbar) 
\end{split} 
\end{equation}
with remainder term
\begin{equation*}
 R_N(t;\hbar) = (\ui \hbar)^{-N} \int_{0}^t  \dots
  \int_{t_{N-1}}^t \hat U(t,t_N) \, \hat U_Q(t_N,0) \, 
  \hat W(t_N) \dots \hat W(t_1) \   
   \ud t_N \dots \ud t_1 \ .
\end{equation*}
In order to estimate the contribution of the remainder when (\ref{eq:Dyson})
is applied to the initial coherent state 
$\psi(0)=\vp^B_{(q,p)}\otimes\phi_{\vecn}$ we use the argument leading to 
(\ref{firstest}) repeatedly. This yields the bound
\begin{equation} 
\label{eq:RemainderEstimate}
 \| R_N(t;\hbar)\psi(0) \| \leq K_N\,\hbar^{N/2} \, t^N \, 
 \theta(t)^{3N}\, \delta(t)^{mN} \ .
\end{equation}
We then replace the symbol of each difference $\hat H - \hat H_Q(t_k)$
appearing in the sum in (\ref{eq:Dyson}) by its Taylor expansion,
\begin{equation}
\label{eq:Taylorremain}
 \sum_{|\nu|=3}^{n_k} \frac{1}{\nu!} H_0^{(\nu)}\bigl(z(t_k)\bigr) 
 \bigl(w-z(t_k)\bigr)^\nu + \hbar \sum_{|\nu|=1}^{n_k-2} \frac{1}{\nu!} 
 \bigl(w-z(t_k)\bigr)^\nu \vecC^{(\nu)}\bigl(z(t_k)\bigr)
 \cdot \ud \pi_s(\vecsigma/2) + r_k(t_k,w) \ .
\end{equation}
The integers $n_k$ are chosen sufficiently large such that, after 
quantisation, the contribution of the remainder $r_k$ to an application 
of (\ref{eq:Dyson}) to $\psi(0)$ can be absorbed in the error estimate 
(\ref{eq:RemainderEstimate}). Similar to the case without spin treated in 
\cite{ComRob97} the quantisation of the main terms in (\ref{eq:Taylorremain}) 
produces matrix valued differential operators 
$\hat p_{kj}(t)=\op^W[p_{kj}(t)]$ with time dependent coefficients acting 
on the coherent state $\vp_{(q,p)}^B\otimes\phi_{\vecn}$. The symbols 
$p_{kj}(t)(x,\xi)$ are polynomials in $(x,\xi)$ of degree $\le k$.
Lemma~\ref{lem:ExpdedWeylOpOnS} finally leads to the following result:
\begin{theorem} 
\label{thm:CsTeFixedS}
Suppose that the quantum Hamiltonian $\hat H$ with symbol (\ref{eq:WeylSymbol})
satisfies the conditions specified in section~\ref{sec1} and the ellipticity
condition (\ref{ellipt1}). Then for $t>0$ and any $N\in\nz$ there exists a 
state $\psi_N(t)\in L^2(\rz^d)\otimes\kz^{2s+1}$, localised at 
$\bigl(q(t),p(t),\vecn(t)\bigr)$, that approximates the full time evolution 
$\psi(t)=\hat U(t,0) \bigl( \varphi_{(q,p)}^B\otimes\phi_{\vecn} \bigr)$ 
of a coherent state up to an error of order $\hbar^{N/2}$. More precisely, 
\begin{equation*}
 \| \psi(t) - \psi_N(t) \| \le C_N  \sum_{j=1}^{N-1} 
 \left( \frac{t}{\hbar} \right)^{j}(\sqrt{\hbar} \theta(t))^{2j+N} \ .
\end{equation*}
The right-hand side vanishes in the combined limits $\hbar\to 0$ and 
$t\to\infty$ as long as $t\ll T_z(\hbar)$, where $T_z(\hbar)$ denotes the
same time scale as in Theorem~\ref{prop:1stapprox}.

Furthermore, $\psi_N(t)$ arises from $\varphi_{(q,p)}^B\otimes\phi_{\vecn}$ 
through the application of certain (time dependent) differential operators 
$\hat p_{kj}(t)=\op^W[p_{kj}(t)]$ of order $\leq k$, followed by the time 
evolution generated by $\hat H_Q(t)$, according to
\begin{equation*}
 \psi_N(t) = \psi_Q(t)+\sum_{(k,j)\in\Delta_N} \hat U_Q(t,0)\,\hat p_{kj}(t)\, 
 \psi(0) \ .
\end{equation*}
Here we have defined $\Delta_N:=\{ (k,j) \in \nz \times
 \nz;\ 1 \le k - 2j \le N-1,\, k \ge 3j, 1 \le j \le N-1 \}$. 
\end{theorem}
We remark that the matrix valued differential operators $\hat p_{kj}(t)$ 
do not increase the frequency set of a semiclassical distribution such as 
the initial state $\varphi_{(q,p)}^B\otimes\phi_{\vecn}$. This follows 
for the translational part from the respective statement without spin 
\cite{Rob87}, whereas the spin part is only acted upon by a matrix producing 
linear combinations of $\vp_{\vecn}$. Moreover, according to 
Proposition~\ref{prop:ExactProp}, $\hat U_Q(t,0)$ propagates the frequency 
set along the trajectory $\bigl(q(t),p(t),\vecn(t)\bigr)$ so that both 
$\psi_Q(t)$ and $\psi_N(t)$ are semiclassically localised at 
$\bigl(q(t),p(t),\vecn(t)\bigr)$.
\subsection{Semiclassics with $\hbar s$ fixed}
We now consider the second semiclassical scenario in which both semiclassical
parameters, $\hbar$ and $s$, are used. For this purpose we still represent
the Hamiltonian $\hat H$ as a matrix valued semiclassical Weyl operator.
That way $\hbar$ appears as before, whereas the second parameter $s\in\nz/2$ 
controls the dimension of the space $\kz^{2s+1}$ on which the symbol operates 
as a linear map. As we will see, the parameter $s$ enters relevant estimates 
through the expression $\hbar\ud\pi_s(\vecsigma/2)$. To leading order this
will produce factors of $\hbar s$. Our desire to perform systematic 
semiclassical expansions therefore forces us to keep the combination
\begin{equation*}
 S: = \hbar s
\end{equation*}
fixed in the semiclassical limit. This means that from now on we consider
$\hbar\to 0$ and $s\to\infty$ with $\hbar s=S$.

An inspection of Proposition \ref{prop:ExactProp} and its proof reveals 
that replacing $\hbar s$ by the constant $S$ will shift the spin-action term
$\vr(t)$, which before was of subleading semiclassical order, to an additional
contribution to the action $R_0$. This suggest that now the translational 
classical dynamics will be influenced by the spin, requiring a modified 
quadratic Hamiltonian. Not only that, revisiting the proof of 
Theorem~\ref{prop:1stapprox} shows that we also have to estimate 
the application of spin operators to spin-coherent states in terms of $s$. 
This requires knowledge of the following:
\begin{lemma}
\label{lem:spinoptocs}
For any $X=\vecx\cdot\vecsigma/2\in\su(2)$, $\vecn\in\Stwo$ and $N\in\nz$ 
there exist differential operators $D^{(j)}_{\vecn}$ of degree $2j$ on 
$C^\infty(\Stwo)\otimes\kz^{2s+1}$ and constants $C_N>0$ such that
\begin{equation} 
\label{eq:RepOpOnSpincs}
 \left \| \ud\pi_s(X)\phi_{\vecn} - \Bigl(s+\frac{1}{2}\Bigr)
 \Bigl(1+\frac{1}{s}\Bigr) \sum_{j=0}^N \frac{1}{s^j} 
 D^{(j)}_{\vecn}\bigl(\vecx\cdot\vecn\,\phi_{\vecn}\bigr) \right \| \le 
 \frac{C_N}{s^{N+1}} \ .
\end{equation}
The leading order in this asymptotic expansion is determined by the
constant $D^{(0)}_{\vecn}=1$,
\begin{equation}
\label{eq:leadinCS}
 \ud\pi_s\bigl(\vecx\cdot\vecsigma/2\bigr)\phi_{\vecn} = 
 s\,\vecx\cdot\vecn\,\phi_{\vecn}\bigl(1+O(s^{-1})\bigr) \ .
\end{equation}
\end{lemma}
\begin{proof}
We start with expressing a linear map $L$ on the representation space 
$\kz^{2s+1}$ in terms of Berezin's quantisation,
\begin{equation}
\label{Berezin}
 L = (2s+1)\int_{\Stwo} P[L](\vecn)\, \Pi(\vecn)\ \ud\vecn \ , 
\end{equation}
where $P[L]$ denotes the upper (or $P$-) symbol of $L$, see e.g. 
\cite{Sim80,Per86}. Furthermore, $\ud\vecn$ is the normalised
area measure on $\Stwo$ and $\Pi(\vecn)$ stands for the projector 
onto the one-dimensional subspace in $\kz^{2s+1}$ spanned by the coherent 
state vector $\phi_{\vecn}$. In the present context the relevant linear maps 
are representation operators of Lie-algebra elements 
$X=\vecx\cdot\vecsigma/2\in\su(2)$. Their upper symbols are simple,
\begin{equation*}
 P[\ud\pi_s(\vecx\cdot\vecsigma/2)](\vecn) = (s+1)\,\vecx\cdot\vecn \ ,
\end{equation*}
see \cite{Sim80,Per86}, so that an application of such an operator to a
coherent state reads
\begin{equation}
\label{coherint}
 \ud\pi_s(\vecx\cdot\vecsigma/2) \phi_{\vecn} = (2s+1)(s+1) \int_{\Stwo} 
 \vecm\cdot\vecx \, \langle \phi_{\vecm}, \phi_{\vecn} \rangle \, 
 \phi_{\vecm} \ \ud\vecm \ .
\end{equation}
The coherent states not being defined globally on $\Stwo$ is irrelevant
to this expression since these states have been defined on a set of 
full measure.

An asymptotic expansion of the integral (\ref{coherint}), as $s\to\infty$, 
can be achieved with the method of steepest descent. This is a variant of 
the stationary phase method, with a complex phase function, and is described 
in detail in \cite{Hoe90}. The first step consists in identifying the
relevant phase factor, which in the present case is given by 
\begin{equation}
\label{eq:steepphase}
 \langle \phi_{\vecn}, \phi_{\vecm} \rangle  = 
 \ue^{\ui s \vp_{\vecn}(\vecm)} \quad\text{with}\quad \im\vp_{\vecn}(\vecm)
 =-\log\left(\frac{1+\vecn\cdot\vecm}{2} \right) \ ,
\end{equation}
where $\vp_{\vecn}$ is independent of $s$, see \cite{Per86}. Outside of a 
neighbourhood of $\vecm =-\vecn$ the function $\im\vp_{\vecn}$ is finite
and non-negative; it has a unique minimum at $\vecm =\vecn$. The real part 
of the phase $\vp_{\vecn}$ can be identified as the area of the spherical 
triangle with edges defined by the north pole, $\vecn$ and $\vecm$. Hence 
$\vecm =\vecn$ is the unique, non-degenerate stationary point of the phase. 
Up to an error of size $O(\ue^{-s})$ one can hence cut out a neighbourhood 
of $\vecm =-\vecn$ from the integral (\ref{coherint}) and use the 
representation (\ref{eq:steepphase}) for $\im\vp_{\vecn}$. The method of 
steepest descent then implies the existence of differential operators 
$D^{(j)}_{\vecn}$ of order $2j$ on $C^\infty(\Stwo)\otimes\kz^{2s+1}$ and 
constants $C_N>0$ such that for any $N\in\nz$ the expansion 
(\ref{eq:RepOpOnSpincs}) holds. The constant $D^{(0)}_{\vecn}$ fixing the 
leading order can be identified by choosing $\vecx=\vecn$, since 
\begin{equation*}
\ud\pi_s\bigl(\vecn\cdot\vecsigma/2\bigr)\phi_{\vecn} = s\phi_{\vecn} \ .
\end{equation*}
Comparing with (\ref{eq:RepOpOnSpincs}) therefore yields $D^{(0)}_{\vecn}=1$, 
which implies (\ref{eq:leadinCS}).
\end{proof}
When constructing a quadratic Hamiltonian we now have to take into account
that an application of a spin operator to a spin-coherent state contributes 
to the leading semiclassical order, as (\ref{eq:leadinCS}) means
\begin{equation*}
 \hat{\vecS}\phi_{\vecn} = S\vecn\,\phi_{\vecn} +O(s^{-1}) \ .
\end{equation*}
We are therefore led to define a quadratic Hamiltonian 
$\hat H_Q(t)=\op^W[H_Q(t)]$ with matrix valued Weyl symbol as follows,
\begin{equation} 
\begin{split}
\label{eq:Hq}
 H_Q(t,w) = &\sum_{|\nu|=0}^2 \frac{1}{\nu!} H_0^{(\nu)}\bigl(z(t)\bigr) 
 \bigl(w-z(t)\bigr)^\nu + S \sum_{|\nu|=1}^2 \frac{1}{\nu!} 
 \vecn(t)\cdot\vecC^{(\nu)}\bigl(z(t)\bigr) \bigl(w-z(t)\bigr)^\nu \\
 &\qquad+ \hbar \vecC\bigl(z(t)\bigr) \cdot \ud\pi_s (\vecsigma/2) \ .
\end{split}
\end{equation}
Like in (\ref{eq:QuadrExpansH}) we have introduced a yet to be determined 
trajectory $z(t)=\bigl(q(t),p(t)\bigr)$ in $\TRd$ with initial condition 
$z(0)=z=(q,p)$, as well as a curve $\vecn(t)$ on $\Stwo$ with 
$\vecn(0)=\vecn$. This Hamiltonian, being quadratic in $(\hat Q,\hat P)$
and linear in $\hat{\vecS}$, propagates an initial coherent state
exactly:
\begin{prop} 
\label{prop:LeadOrdProp}
The solution of the quadratic Schr\"odinger equation
\begin{equation}
\label{eq:QuadratSch}
 \ui\hbar\frac{\partial\psi_Q}{\partial t}(t) = \hat H_Q(t)\psi_Q(t)
 \quad\text{with}\quad \psi_Q(0)= \varphi_{(q,p)}^B \otimes \phi_{\vecn}
\end{equation}
is, up to an additional phase, again a coherent state,
\begin{equation}
\label{eq:psiQansatz}
 \psi_Q(t) =  \ue^{\ui\left( \frac{R_{\mathrm{so}}(t)}{\hbar} + 
 \frac{\pi}{2}\sigma(t) \right)} \varphi_{(q(t),p(t))}^{B(t)} \otimes
 \phi_{\vecn(t)} \ .
\end{equation}
Here $\bigl(q(t),p(t),\vecn(t)\bigr)=\Phi^t_{\mathrm{so}}(p,q,\vecn)$ is
the solution of Hamilton's equations of motion (\ref{SOflow}) on 
$\TRd\times\Stwo$ generated by the classical spin-orbit Hamiltonian 
\begin{equation}
\label{eq:soHamil}
H_{\mathrm{so}}(x,\xi,\vecn) := H_0(x,\xi)+ S\vecn\cdot\vecC(x,\xi) \ .
\end{equation}
The phase of $\psi_Q(t)$ is determined by 
\begin{equation}
\label{eq:Rso}
 R_{\mathrm{so}}(t)= \int_0^t \bigl(p(t')\dot q(t') - 
 H_0(q(t'),p(t'))\bigr) \ \ud t' +S\vr(t)\ ,
\end{equation}
which can be viewed as a total spin-orbit principal function, and by the 
Maslov phase $\si(t)$. The latter derives from the time evolution
\begin{equation}
\label{eq:B(t)2}
 B(t) = \bigl(\dpr_q q(t) B + \dpr_p q(t)\bigr) \bigl(\dpr_q p(t) B 
 + \dpr_p p(t)\bigr)^{-1}
\end{equation}
of the complex symmetric $d\times d$ matrix $B\in\Sigma_d$.
\end{prop}
\begin{proof}
The proof of this proposition parallels that of 
Proposition~\ref{prop:ExactProp}; however, a few modifications are necessary. 
One can again consider (\ref{eq:psiQansatz}) as an ansatz and determine its 
ingredients by inserting it into (\ref{eq:QuadratSch}), leading to equations 
analogous to (\ref{transeval}). As opposed to (\ref{conditions}) the fact 
that now $S=\hbar s$ is fixed shifts the term with $\vr(t)$ from the last 
equation to the first one. Moreover, due to the modified definition of the 
quadratic Hamiltonian the principal symbol $H_0$ is replaced by 
$H_{\mathrm{so}}$ in all places but one, yielding
\begin{equation*} 
\begin{split}
 \dot \theta 
      & = \dot q p -H_0 +  S \dot \vr  \\
 - \dot p + B \dot q 
      &= H'_{\textrm{so},x} + B H'_{\textrm{so},\xi} \\
 - \dot B 
      &= H''_{\textrm{so},xx} + H''_{\textrm{so},\xi x} B
         + B H''_{\textrm{so},\xi x} + B H''_{\textrm{so},\xi \xi} B \\
 \frac{\dot\gamma}{\gamma} 
      &= - \frac{1}{2}\mtr\bigl( H''_{\textrm{so},\xi x} + 
         H''_{\textrm{so},\xi \xi}B\bigr) \ .
\end{split} 
\end{equation*}
The first two equations fix the translational part of the classical dynamics
to be solutions of (\ref{SOflow}) with some $\vecn(t)$ and yield the
spin-orbit principal function (\ref{eq:Rso}). In the last two equations 
$H_{\textrm{so}}$, which is evaluated at $\bigl(q(t),p(t),\vecn(t)\bigr)$,
can be viewed as a time dependent Hamiltonian, $\tilde{H}_{\textrm{so}}(w,t)
=H_{\textrm{so}}(w,\vecn(t))$, for the translational degrees of freedom, with 
the time dependence introduced through $\vecn(t)$. These equations can be 
solved in the same manner as in the time independent case, yielding 
\begin{equation*}
 B(t) = S_{\textrm{so},z}(t)[B] 
\end{equation*}
as in (\ref{SofB}). Here $S_{\textrm{so},z}(t)$ is a solution of
\begin{equation}
\label{eq:Sspinorbitdef}
 \frac{\ud}{\ud t}S_{\textrm{so},z}(t) = 
 J\tilde{H}''_{\textrm{so}}\bigl(z(t),t\bigr)\,S_{\textrm{so},z}(t)
\end{equation}
with $S_{\textrm{so},z}(0)=\eins_{2d}$; it hence yields (\ref{eq:B(t)2}).

The classical spin motion $\vecn(t)$ so far has remained undetermined. Since
the equation for the spin-coherent state is again (\ref{spinderivative}),
it follows that $\vecn(t)$ must be a solution to the spin part of
(\ref{SOflow}).
\end{proof}
In contrast to the previous case the classical dynamics that governs the 
time evolution of the coherent state $\psi_Q(t)$ now is Hamiltonian on the 
product phase space $\TRd\times\Stwo$, see (\ref{SOflow}). This means that 
the spin precession is not merely following the translational motion, but 
there occurs a mutual influence of both dynamics. This effect is caused by 
the energy scales of the translational and the spin dynamics being comparable 
in the semiclassical limit, whereas when $s$ is fixed the energy scale of 
the translational motion dominates. 

We now compare the time evolution generated by the full Hamiltonian $\hat H$ 
with the approximate dynamics following from the quadratic Hamiltonian 
$\hat H_Q(t)$ whose symbol is given in (\ref{eq:Hq}). As opposed to the 
situation discussed previously, see (\ref{ellipt1}), when keeping $\hbar s$ 
fixed the ellipticity condition has to be imposed on the full symbol of 
$\hat H$, see (\ref{ellipt}), which implies
\begin{equation*}
 c\,\bigl( 1+|x|^2\bigr)^{-M_x/2}\bigl(1+|\xi|^2 \bigr)^{-M_\xi/2} \geq
 \bigl\|\bigl(H(x,\xi)+\ui\bigr)^{-1}\bigr\| \geq 
 \frac{\bigl\|\bigl(H(x,\xi)+\ui\bigr)^{-1}\psi\bigr\|}{\|\psi\|} \ .
\end{equation*}
Here in the middle $\|\cdot\|$ denotes the operator norm on $\kz^{2s+1}$,
and on the right-hand side $\psi$ is any non-zero vector in $\kz^{2s+1}$. 
Choosing $\psi = \bigl(H(x,\xi)+\ui\bigr)^2\phi_{\vecn}$ and using 
(\ref{eq:leadinCS}) we then conclude that the spin-orbit Hamiltonian 
(\ref{eq:soHamil}) is elliptic, in the sense that 
\begin{equation*}
 |H_{\mathrm{so}}(x,\xi,\vecn)+\ui| \geq C\,
 \bigl( 1+|x|^2\bigr)^{M_x/2}\bigl(1+|\xi|^2 \bigr)^{M_\xi/2} 
\end{equation*}
holds for all $(x,\xi,\vecn)\in\TRd\times\Stwo$. Therefore, we can again base 
our further investigation of the difference between the two quantum dynamics 
on the Duhamel relation (\ref{Duhamel}). This requires to estimate the 
analogue of (\ref{eq:DiffHamiltonians}), where in the present situation 
$\hat H-\hat H_Q(t)$ is the Weyl quantisation of the symbol
\begin{equation} 
\label{eq:DiffH_HQ}
\begin{split}
 \bigl( H-H_Q(t)\bigr)(w) 
 &= \sum_{|\nu|=1}^2\frac{1}{\nu!} \bigl( \hbar\ud\pi_s(\vecsigma/2) 
    -S\vecn(t)\bigr)\cdot\vecC^{(\nu)}\bigl(z(t)\bigr)\bigl(w-z(t)\bigr)^\nu\\
 &\qquad + H_0^{[3]}(t,w) + \hbar \vecC^{[3]}(t,w)\cdot
     \ud\pi_s(\vecsigma/2) \ ,
\end{split}
\end{equation}
in which $H_0^{[3]}$ and $\vecC^{[3]}$ denote Taylor remainders of order 
three. Introducing an operator $\hat W(t)$ as in (\ref{Wdef}), the same type 
of an Egorov theorem as above applies, leading to the symbol
\begin{equation}
\label{eq:QEgorov1}
 W(t,w) = d^\ast\bigl(z(t)\bigr) \, \bigl( H-H_Q(t)\bigr)
 \bigl(z-S_{\mathrm{so},z}^{-1}(t)(w-z(t))\bigr) \, d\bigl(z(t)\bigr) 
\end{equation}
of $\hat W(t)$. We remark that $z(t)$ being the projection of 
$\Phi^t_{\mathrm{so}}(z,\vecn)$ to $\TRd$ here requires the differential 
$S_{\mathrm{so},z}(t)$ of $\Phi^t_{\mathrm{so}}$ with respect to $z$. The 
conjugation with $d\bigl(z(t)\bigr)$ has no effect on the scalar terms in 
(\ref{eq:DiffH_HQ}), whereas it rotates the spin operator to
$R\bigl(g(t)\bigr)\hbar\ud\pi_s(\vecsigma/2)$. Hence, for the application
of (\ref{eq:QEgorov1}) to a spin-coherent state $\phi_{\vecn}$ we can 
employ Lemma~\ref{lem:spinoptocs}. By also converting estimates with respect
to $s$ into ones with respect to $\hbar$ this yields to leading order
\begin{equation}
\label{eq:applytoCS}
 \left(R\bigl(g(t)\bigr)\hbar\ud\pi_s(\vecsigma/2)-S\vecn(t)\right)
 \phi_{\vecn} = S\left(R\bigl(g(t)\right)\vecn - \vecn(t)\bigr)\phi_{\vecn}
 +O(s^{-1}) = O(\hbar) \ .
\end{equation}
Moreover, the complete asymptotic series in powers of $s^{-1}$ provided by 
Lemma~\ref{lem:spinoptocs} results in a full asymptotic expansion of
(\ref{eq:applytoCS}) in powers of $\hbar$. This observation now enables us
to apply Lemma~\ref{lem:ExpdedWeylOpOnS} in a completely analogous way to 
that used previously, yielding
\begin{equation*}
 \| (\hat H - \hat H_Q(t)) \psi_Q(t) \| \leq K \hbar^{3/2} 
 \theta(t)^3 \, \delta(t)^m \ .
\end{equation*}
Here the quantities $\theta(t)$ and $\delta(t)$ are defined as in 
(\ref{thetdeldef}), however, now with the differential $S_{\mathrm{so},z}(t)$,
and $z(t)$ as given in Proposition~\ref{prop:LeadOrdProp}. 

The stability of the trajectory $z(t)$ is encoded in the quantity
\begin{equation}
\label{eq:Sasymptso}
 \tilde{\la}_{{\mathrm{max}}} (z) = \limsup_{t\to\infty} \frac{1}{t}
 \log\|S_{\mathrm{so},z}(t)\|_{\mathrm{HS}} \ .
\end{equation}
Since $z(t)$ is not the integral curve of a flow, rather than calling
$\tilde{\la}_{{\mathrm{max}}} (z)$ a Lyapunov exponent we refer to it as a
stability exponent. This can, however, be bounded by the maximal Lyapunov
exponent of the flow-line $(z(t),\vecn(t))$ in $\TRd\times\Stwo$, see
the appendix. Thus, in close analogy to Theorem~\ref{prop:1stapprox} we 
finally obtain:
\begin{theorem}
\label{prop:1stapprox2}
Let the conditions imposed on the Hamiltonian in section~\ref{sec1} hold. 
Then the coherent state $\psi_Q(t)$ defined in (\ref{eq:psiQansatz}) 
semiclassically approximates 
$\psi(t)=\hat U(t,0) \bigl( \vp_{(q,p)}^B\otimes\phi_{\vecn} \bigr)$ 
in the following sense,
\begin{equation*}
 \| \psi(t) - \psi_Q(t) \| \le K\sqrt{\hbar}\,t\,\theta(t)^3 \  ,
\end{equation*}
when $\hbar s$ is kept fixed. The right-hand side vanishes in the combined 
limits $\hbar\to 0$, $s\to\infty$ and $t\to\infty$ as long as 
$t\ll T_z(\hbar)$. The time scale $T_z(\hbar)$ depends on the linear stability
of the trajectory $z(t)$. If the latter possesses a positive and finite 
stability exponent $\tilde{\la}_{{\mathrm{max}}}(z)$, one has 
$T_z(\hbar)=\frac{1}{6\tilde{\la}_{{\mathrm{max}}}(z)}|\log\hbar|$. 
In case $z(t)$ is a projection to $\TRd$ of a trajectory $(z(t),\vecn(t))$ 
on a (non-degenerate) KAM-torus in $\TRd\times\Stwo$ this time scale is 
$T_z(\hbar)=C\,\hbar^{-1/8}$.
\end{theorem}
As in the previous case an improvement of the semiclassical error can be 
achieved with the Dyson expansion (\ref{eq:Dyson}). The present case, however,
requires an additional estimate of the spin contribution in terms of $s$.
Concerning the error term $R_N(t;\hbar)\psi(0)$, the translational part
is dealt with by a repeated application of the argument leading to 
Theorem~\ref{prop:1stapprox2}. For the spin part an inspection of the 
relations (\ref{eq:DiffH_HQ}) and (\ref{eq:QEgorov1}) reveals the necessity 
to estimate the successive application of the operators
\begin{equation*} 
 \Lambda(t_k):= \vecC^{(\nu)}\bigl(z(t_k)\bigr)\cdot 
 \Bigl( R\bigl(g(t_k)\bigr)\hbar \ud \pi_s(\vecsigma/2) - S \vecn(t_k)\Bigr)
\end{equation*} 
to the spin-coherent state $\phi_{\vecn}$. Representing these operators in 
the form (\ref{Berezin}), the result of their $l$-fold ($l\leq j$) 
application reads
\begin{equation} 
\label{eq:MultipleRepOp}
\begin{split}
 \Lambda(t_l) \dots \Lambda(t_1) \phi_{\vecn} = (2s+1)^l \int\limits_{\Stwo}
 \dots \int\limits_{\Stwo}  
 &P[\Lambda(t_l)](\vecm_l) \dots P[\Lambda(t_1)](\vecm_1) \times \\ 
 &\times \Pi(\vecm_l) \dots \Pi(\vecm_1)\phi_{\vecn} \  \ud\vecm_l\dots
   \ud\vecm_1 \ ,
\end{split}
\end{equation}
with the lower symbols
\begin{equation}
\label{eq:prodsymbol}
 P[\Lambda(t_k)](\vecm_k) = \vecC^{(\nu)}\bigl(z(t_k)\bigr)\cdot
 \Bigl( S \bigl( R(g(t_k)\bigr) \vecm_{k}-\vecn(t_k)\bigr) + 
 \hbar R\bigl(g(t_k)\bigr) \vecm_{k} \Bigr) \ .
\end{equation}
Starting with $\vecm_l$, the integral (\ref{eq:MultipleRepOp}) can be 
successively evaluated with the method of steepest descent similar to 
the proof of Lemma~\ref{lem:spinoptocs}. The relation
\begin{equation*}
 \Pi(\vecm_l) \dots \Pi(\vecm_1) \phi_{\vecn} = 
 \langle \phi_{\vecm_l}, \phi_{\vecm_{l-1}} \rangle \cdots \langle 
 \phi_{\vecm_1}, \phi_{\vecn} \rangle \phi_{\vecm_j}
\end{equation*}
then shows that the critical points of the phase are given by 
$\vecm_l=\vecm_{l-1}=\cdots=\vecm_1=\vecn$. At these points, however, the 
lower symbols $P[\Lambda(t_k)](\vecm_k)$ are of order $\hbar$, compare
(\ref{eq:prodsymbol}). The application of the method of steepest descent 
therefore yields in leading order a contribution $O(\hbar^l)=O(s^{-l})$. 
Derivatives of total order $n$ contribute terms of the order 
$O(s^{-n}\hbar^{l-n})=O(s^{-l})$, if $n\leq l$, and of the order $O(s^{-n})$ 
otherwise. Altogether there hence exist differential operators 
$\cD^{(\kappa)}$ of order $\leq 2\kappa$ on 
$C^\infty((\Stwo)^l)\otimes\kz^{2s+1}$ such that
\begin{equation} 
\label{eq:ExpandMultRepOp}
 \Lambda(t_l) \dots \Lambda(t_1) \phi_{\vecn} -
 \Bigl(1+\frac{1}{2s}\Bigr)^l \sum_{\kappa=l}^K \frac{1}{s^{\kappa}} 
 \cD^{(\kappa)} \bigl(P[\Lambda(t_l)](\vecm_l) \dots P[\Lambda(t_1)](\vecm_1) 
 \phi_{\vecn} \bigr)_{\vecm_l=\cdots=\vecm_1=\vecn} 
\end{equation}
is of the order $s^{-(K+1)}$ for any $K \ge l$. The left-hand side of 
(\ref{eq:MultipleRepOp}) hence is of the order $O(s^{-l})=O(\hbar^l)$,
meaning that every factor $\Lambda(t_k)$ contributes a factor of $\hbar$.
We therefore finally obtain an estimate of the remainder term to the Dyson 
series given by
\begin{equation*}
 \| R_N(t;\hbar) \psi(0) \| \le K_N \hbar^{N/2} t^N \theta(t)^{3N}
 \delta(t)^{mN} \ .
\end{equation*}
The main terms in the Dyson expansion are treated by replacing each factor
of (\ref{eq:DiffH_HQ}), occurring at $t=t_k$, with the Taylor expansions 
\begin{equation*} 
\begin{split}
 \sum_{|\nu|=1}^2 
 &\frac{1}{\nu!} \bigl( \hbar\ud\pi_s(\vecsigma/2) -S\vecn(t_k)\bigr)\cdot
    \vecC^{(\nu)}\bigl(z(t_k)\bigr)\bigl(w-z(t_k)\bigr)^\nu\\
 &+\sum_{\nu=3}^{n_k}\frac{1}{\nu!}\Bigl( H_0^{(\nu)}\bigl(z(t_k)\bigr)
    +\hbar\vecC^{(\nu)}\bigl(z(t_k)\bigr)\ud\pi_s(\vecsigma/2)\Bigr) 
    \bigl(w-z(t_k)\bigr)^\nu + r_k(t_k,w)\ ,
\end{split}
\end{equation*}
where again the integers $n_k$ are chosen sufficiently large. The contribution
of the translational degrees of freedom can be dealt with as in the previous
semiclassical scenario, and the spin contribution follows from the expansion
(\ref{eq:ExpandMultRepOp}). Finally grouping together terms of corresponding 
orders in $\hbar$, we arrive at a statement analogous to 
Theorem~\ref{thm:CsTeFixedS}.
\begin{theorem} 
\label{thm:CsTeFixedhbarS}
Suppose that the quantum Hamiltonian $\hat H$ with symbol (\ref{eq:WeylSymbol})
satisfies the conditions specified in section~\ref{sec1}. Then for $t>0$ and 
any $N\in\nz$ there exists a state $\psi_N(t)\in L^2(\rz^d)\otimes\kz^{2s+1}$, 
localised at $\bigl(q(t),p(t),\vecn(t)\bigr)$, that approximates the full time 
evolution 
$\psi(t)=\hat U(t,0) \bigl( \varphi_{(q,p)}^B\otimes\phi_{\vecn} \bigr)$ 
of a coherent state up to an error of order $\hbar^{N/2}$ when $\hbar s$ 
is fixed. More precisely, 
\begin{equation*}
 \| \psi(t) - \psi_N(t) \| \le C_N  \sum_{j=1}^{N-1} 
 \left( \frac{t}{\hbar} \right)^{j}(\sqrt{\hbar} \theta(t))^{2j+N} \ .
\end{equation*}
The right-hand side vanishes in the combined limits $\hbar\to 0$, 
$s\to\infty$ and $t\to\infty$ as long as $t\ll T_z(\hbar)$, where $T_z(\hbar)$
denotes the same time scale as in Theorem~\ref{prop:1stapprox2}.

Furthermore, $\psi_N(t)$ arises from $\varphi_{(q,p)}^B\otimes\phi_{\vecn}$ 
through the application of certain (time dependent) differential operators 
$\hat q_{k\kappa j}(t)=\op^W[p_{kj}(t)]\otimes r_\kappa$, 
\begin{equation*}
 \psi_N(t) = \psi_Q(t)+\sum_{(k,\kappa,j)\in\Delta_N} \hat U_Q(t,0)\,
 \hat q_{k\kappa j}(t)\, \psi(0) \ ,
\end{equation*}
where $p_{kj}(t)$ is a polynomial in $(x,\xi)$ of degree $\leq k$ and 
$r_\kappa$ is a differential operator of order $\leq 2\kappa$ on 
$C^\infty(\Stwo)\otimes\kz^{2s+1}$. Here we have also defined 
\begin{equation*} 
 \Delta_N := \{ (k,\kappa,j) \in \nz^3; \ 1 \le k + 2 \kappa - 2j \le N-1,\, 
 k + 2 \kappa \ge 3j, \, 1 \le j \le N-1\} \ .
\end{equation*}
\end{theorem}
The semiclassical localisation of $\psi_N(t)$ here is different from the
situation covered by Theorem~\ref{thm:CsTeFixedS} in that the operators 
$r_\kappa$ act on $\phi_{\vecn}$. But these are differential operators
and hence do not increase the frequency set. This means that $\psi_N(t)$ 
is semiclassically localised at $\Phi^t_{\mathrm{so}}(q,p,\vecn)$ and in 
in this respect is not different from the classically propagated coherent
state $\psi_Q(t)$.


%% file: 4sec.tex
\section{Discussion}
\label{sec4}
In the previous section we analysed the semiclassical behaviour of coherent 
states in two different limits. In various places we saw that the difference 
between the two cases is expressed in the way the classical translational 
and spin motion are coupled. Otherwise the final results agree to a large 
extent. This includes the mechanisms of semiclassical localisation in the 
product phase space $\TRd\times\Stwo$. 

The problem of how the localisation of an initial coherent state develops 
with time can be made more explicit by using semiclassical phase-space lifts 
of the coherent states. At $t=0$ the state 
$\psi(0)=\vp_{(q,p)}^B\otimes\phi_{\vecn}$ is concentrated in a neighbourhood 
of the point $(q,p,\vecn)\in\TRd\times\Stwo$. This concentration can be 
measured in terms of expectation values $\langle\psi(0),\hat A\psi(0)\rangle$ 
of operators $\hat A=\op^W[A]$ that are quantisations of well localised 
symbols $A\in C^\infty_0(\TRd)\otimes{\mathrm{M}}_{2s+1}(\kz)$. For simplicity
we also assume that $A$ is independent of $\hbar$. At later times $\psi(t)$ 
can in both semiclassical scenarios be approximated by an appropriate
coherent state $\psi_Q(t)$, such that
\begin{equation}
\label{localiset}
 \langle\psi(t),\hat A\psi(t)\rangle = 
 \langle\psi_Q(t),\hat A\psi_Q(t)\rangle + o(1) \ , \quad t\ll T_z(\hbar)\ .
\end{equation}
The expectation value on the right-hand side has a phase-space representation
\begin{equation}
\label{expectPS}
 \langle\psi_Q(t),\hat A\psi_Q(t)\rangle = \frac{1}{(2\pi\hbar)^d}\iint_{\TRd}
 W[\vp^{B(t)}_{z(t)}](w)\,\langle\phi_{\vecn(t)},A(w)\phi_{\vecn(t)}
 \rangle_{\kz^{2s+1}} \ \ud w \ .
\end{equation}
A comparison with (\ref{Wignercohvec}) clearly reveals that the state 
$\psi_Q(t)$ is concentrated at the point $(q(t),p(t),\vecn(t))$ in the 
semiclassical limit as long as the quadratic form $G_{B(t)}/\hbar$ is 
strictly positive definite. Either of the time evolutions (\ref{Bsolve}) 
and (\ref{eq:B(t)2}) of $B$ now imply \cite{Sch01}
\begin{equation*}
 G_{B(t)} = (S_{z}(t)^{-1})^* \, G_B\, S_{z}(t)^{-1} \ ,
\end{equation*}
so that the spreading of $\psi_Q(t)$ in $\TRd$, see (\ref{eq:variance}), 
is bounded according to
\begin{equation*}
 \frac{\hbar}{\mtr G_{B(t)}} \leq 
 \frac{\hbar}{\|G_{B(t)}\|_{\mathrm{HS}}} \leq 
 \frac{\hbar\|S_{z}(t)\|_{\mathrm{HS}}^2}{\|G_B\|_{\mathrm{HS}}} \ .
\end{equation*}
Here $S_{z}(t)$ denotes the differential of the appropriate flow with
respect to $(x,\xi)$. If $z(t)$ now is a trajectory with maximal Lyapunov 
(or stability) exponent $\la_{{\mathrm{max}}}(z)>0$, the requirement for 
the state $\psi_Q(t)$ to remain localised therefore is 
$t\ll\frac{1}{2\la_{\mathrm{max}}(z)}|\log\hbar|$. This time scale is 
three times larger than $T_z(\hbar)$, which is the estimated time in 
(\ref{localiset}) for the coherent state $\psi_Q(t)$ to still well 
approximate the full time evolution $\psi(t)$. 

Let us remark that the limitations in (\ref{localiset}), to approximate
the expectation value in terms of a coherent state, derive from estimating
the difference $\psi(t)-\psi_Q(t)$ in $L^2$-norm. But the error term on 
the right-hand side of (\ref{localiset}) measures this difference in a 
considerably weaker form so that one might expect it to vanish as $\hbar\to 0$
and $t\to\infty$ also for times $T_z(\hbar)\leq t\ll 3T_z(\hbar)$. In the
case without spin Bouzouina and Robert \cite{BouRob02} proved that this 
indeed holds, suggesting that the same is true in the present setting.

Expectation values in coherent states such as (\ref{localiset}) can also be 
used to obtain the leading semiclassical description of the propagation of
observables. To see this let $\hat A$, as above, be a bounded Weyl 
operator and denote its quantum time evolution by 
$\hat A(t)=\hat U(t,0)^\ast\hat A\,\hat U(t,0)$. Here, however, 
we do not necessarily require the symbol to be compactly supported. The
relations \eqref{localiset} and \eqref{expectPS} then remain valid so that
\begin{equation*}
 \langle \psi(0), \hat A(t) \psi(0) \rangle = \frac{1}{(2 \pi \hbar)^d} 
 \iint_{\TRd} W[\varphi_{z(t)}^{B(t)}](w) \, \langle \phi_{\vecn(t)}, A(w)
 \phi_{\vecn(t)} \rangle_{\kz^{2s+1}} \ \ud w + o(1) \ .
\end{equation*}
Since $\hat A(t)$ is bounded it may also be expressed as a Weyl operator, 
with symbol $A(t)$ such that for $t\ll T_z(\hbar)$ equation \eqref{localiset} 
can be rewritten as
\begin{equation*} 
\begin{split} 
 \frac{1}{(2 \pi \hbar)^d} & \iint_{\TRd}
  W[\varphi_{z(0)}^{B(0)}](w)\,\langle \phi_{\vecn(0)}, A(t)(w)\phi_{\vecn(0)} 
 \rangle_{\kz^{2s+1}} \ \ud w \\ 
 &- \frac{1}{(2 \pi \hbar)^d} \iint_{\TRd} W[\varphi_{z(t)}^{B(t)}](w) \, 
 \langle \phi_{\vecn(t)}, A(w) \phi_{\vecn(t)}\rangle_{\kz^{2s+1}} \ \ud w 
 = o(1) \ .
\end{split} 
\end{equation*}
The semiclassical localisation properties of the coherent states discussed 
above therefore imply that in leading order the symbol of the time evolved 
observable $\hat A(t)$ can be expressed in terms of the symbol of $\hat A$ 
transported along the classical flow $\bigl(q(t),p(t),\vecn(t)\bigr)$, 
\begin{equation*} 
 \langle \phi_{\vecn}, A(t)(q,p) \phi_{\vecn} \rangle_{\kz^{2s+1}}
 - \langle \phi_{\vecn(t)},A\bigl(q(t),p(t)\bigr) \phi_{\vecn(t)} 
 \rangle_{\kz^{2s+1}} = o(1)\ .
\end{equation*} 
The $\kz^{2s+1}$-expectation values in spin-coherent states are lower (or $Q$-)
symbols (see e.g. \cite{Sim80,Per86}) of the matrix valued functions $A(t)$ 
and $A$, respectively. In terms of this mixed phase space representation of 
operators, employing Weyl calculus for the translational part and $Q$-symbols 
for the spin part, this means that the quantum time evolution of observables 
follows the classical dynamics in leading semiclassical order. This statement
represents a limited version of an Egorov theorem and again is valid for
both semiclassical scenarios discussed in the preceding section, up to the 
time scale $t\ll T_z(\hbar)$.

%% file: app.stability.tex
\label{app.stability}
The flows $\Phi^t_0$ and $\Phi^t_{so}$ introduced in section \ref{sec1}
are both Hamiltonian flows on symplectic phase spaces. They are generated
by smooth Hamiltonian functions $H$ on $2n$-dimensional smooth manifolds 
$M$ with symplectic forms $\om$. In the first case the Hamiltonian is 
$H_0(x,\xi)$, defined on the phase space $M=\TRd$ so that $n=d$ and 
$\om=\ud x\wedge\ud\xi$. In the situation of classical spin-orbit coupling
the Hamiltonian $H_{\mathrm{so}}(x,\xi,\vecn)$ is given on $M=\TRd\times\Sph$.
Thus, $n=d+1$ and $\om=\ud x\wedge\ud\xi +\ud\vecn$, where $\ud\vecn$
denotes the normalised area two-form on the sphere $\Sph$. In this appendix 
we want to recall the notion of Lyapunov exponents and give sufficient 
criteria of their existence in terms of properties of the Hamiltonian 
function. 

The linear stability of a flow $\Phi^t$ is determined by properties of 
the differential $\uD\Phi^t(\al)$ which is a linear map from the tangent 
space ${\mathrm{T}}_\al M$ to ${\mathrm{T}}_{\Phi^t(\al)}M$. It, moreover, 
is a multiplicative cocycle over the flow $\Phi^t$, i.e. 
$\uD\Phi^{t+t'}(\al)=\uD\Phi^{t'}(\Phi^t(\al))\,\uD\Phi^t(\al)$. If one 
introduces a euclidean scalar product in the tangent spaces, this gives rise 
to the adjoint $\uD\Phi^t(\al)^*$. Then $\uD\Phi^t(\al)^*\,\uD\Phi^t(\al)$ 
is a non-negative symmetric linear map on ${\mathrm{T}}_\al M$ whose 
eigenvalues we denote by 
\begin{equation*}
 \mu_t^{(1)}(\al) \geq\dots\geq \mu_t^{(2n)}(\al) \geq 0\ .
\end{equation*}
The $2n$ Lyapunov exponents of the flow $\Phi^t$ at $\al\in M$ are now
given by the expressions
\begin{equation*}
 \la_k (z) := \limsup_{t\to\infty} \frac{1}{2t}\log\mu_t^{(k)}(z)\ , 
\end{equation*}
if these are finite. The largest Lyapunov exponent $\la_{\mathrm{max}}(\al)$ 
provides a quantitative measure for the linear stability of $\Phi^t$ since 
it measures the leading rate of local phase space expansion; it can be 
obtained from the relation
\begin{equation}
\label{MaxLyapunov}
 \la_{\mathrm{max}} (\al) = \limsup_{t\to\infty} \frac{1}{2t}
 \log\mtr \bigl( \uD\Phi^t(\al)^*\,\uD\Phi^t(\al) \bigr) \ .
\end{equation}
Hamiltonian flows leave the energy shells 
\begin{equation*}
 \Om_E := \{ \al\in M;\ H(\al)=E \}
\end{equation*}
invariant. If $E$ is a regular value of the Hamiltonian function $H$,
the energy shell $\Om_E$ is a smooth submanifold of $M$ of dimension 
$2n-1$. In such a case two Lyapunov exponents are always zero. They 
correspond to the direction of the flow and the direction transversal to 
the energy shell. Of the remaining $2n-2$ Lyapunov exponents half are
non-negative (if they exist) and the rest of the Lyapunov spectrum is given
by minus the first half.

In general it is not known whether the Lyapunov exponents are finite. If,
however, an energy shell $\Om_E$ is compact, one can introduce the normalised
Liouville measure as a flow invariant probability measure on $\Om_E$. In 
this case one can apply Oseledec' multiplicative ergodic theorem to
the restriction of $\Phi^t$ to this energy shell \cite{Ose68}; it guarantees 
that the Lyapunov exponents are finite for almost all points on $\Om_E$ with
respect to Liouville measure. Moreover, if the flow is ergodic with respect
to Liouville measure $\la_k(\al)$ is constant on a set of full measure.  
Since we want to consider also non-compact energy shells we now give 
alternative sufficient criteria for the finiteness of Lyapunov spectra.
\begin{prop} 
\label{prop:Lyapfinite}
Let $H\in C^\infty(M)$ be a Hamiltonian function such that the 
Hilbert-Schmidt norm of $\uD^2 H$ is bounded on the energy shell $\Om_{E,\al}$
that contains the point $\al\in M$. Then the Lyapunov exponents 
$\la_1(\al),\dots,\la_{2n}(\al)$ are finite. 
\end{prop}
\begin{proof}
Fix $\al\in M$ and introduce canonical coordinates 
$(q,p)\in U\subset\rz^n\times\rz^n$ in a neighbourhood of $\al$. Then in this
neighbourhood $\uD^2 H$ is represented by the matrix $H''(q,p)$ of second 
derivatives with respect to $(q,p)$. In these coordinates we denote the flow 
by $\tilde\Phi^t(q,p)$; its differential satisfies the equation
\begin{equation}
\label{cocycleDGL}
 \frac{\ud}{\ud t}\uD\tilde\Phi^t(q,p) = J\,H''\bigl(\tilde\Phi^t(q,p)\bigr)\,
 \uD\tilde\Phi^t(q,p) \ , \qquad \uD\tilde\Phi^t(q,p)|_{t=0} =\eins_{2n}\ ,
\end{equation}
where $J=\left(\begin{smallmatrix} 0&\eins_n \\ -\eins_n &0 \end{smallmatrix}
\right)$. By integrating (\ref{cocycleDGL}) and taking the Hilbert-Schmidt 
norm one obtains
\begin{equation*}
 \| \uD\tilde\Phi^t(q,p) \|_{\mathrm{HS}} \leq 2n + \int_0^t 
 \|J\,H''\bigl(\tilde\Phi^s(q,p)\bigr)\|_{\mathrm{HS}} \, 
 \|\uD\tilde\Phi^s(q,p)\|_{\mathrm{HS}} \ \ud s \ .
\end{equation*}
For simplicity we here assume that for the points $\Phi^s(\al)$, $s\in [0,t]$,
one can use the same system of canonical coordinates. Gronwall's inequality 
then yields the estimate $(t>0)$
\begin{equation*}
 \| \uD\tilde\Phi^t(q,p) \|_{\mathrm{HS}} \leq 2n 
 \exp\Bigr\{ t\sup_{s\in [0,t]} 
 \|J\,H''\bigl(\tilde\Phi^s(q,p)\bigr)\|_{\mathrm{HS}} \Bigr\}
 \leq 2n\,\ue^{Ct} \ ,
\end{equation*}
with some constant $C>0$. The last line follows from the boundedness of 
$\uD^2 H$ on $\Om_{E,\al}$. Since on the other hand 
\begin{equation*}
 \| \uD\tilde\Phi^t(q,p) \|_{\mathrm{HS}} = 
 \sqrt{\mu_t^{(1)}(\al) + \dots + \mu_t^{(2n)}(\al)} \ ,
\end{equation*}
the bound 
\begin{equation*}
 \frac{1}{2t}\log\mu_t^{{\mathrm{max}}}(\al) \leq K
\end{equation*}
for the maximal eigenvalue $\mu_t^{\mathrm{max}}(\al)$ follows. This finally 
implies the assertion.
\end{proof}
An application of this Proposition to the two flows $\Phi^t_0$ (defined
on $M=\TRd$) and $\Phi^t_{so}$ (defined on $M=\TRd\times\Sph$) immediately
yields
\begin{cor}
\label{cor:Lyapfinite}
If the norm of $H_0''$ is bounded on $\Om_{E,(x,\xi)}\subset\TRd$, the 
$2d$ Lyapunov exponents $\la_{0,k}(x,\xi)$ of the flow $\Phi_0^t$ are 
finite. If, in addition, the derivatives $\vecC^{(\nu)} (x',\xi')$ of 
order $|\nu|\leq 2$ are bounded for all 
$ (x',\xi',\vecn')\in\Om_{E,(x,\xi,\vecn)}\subset\TRd\times\Sph$, 
the $2d+2$ Lyapunov exponents $\la_{\mathrm{so},k}(x,\xi,\vecn)$ of the flow 
$\Phi_{\mathrm{so}}^t$ are also finite.
\end{cor}
In the second semiclassical scenario, however, rather than the Lyapunov
exponent $\la_{\mathrm{so},k}(q,p,\vecn)$ of a point 
$(q,p,\vecn)\in\TRd\times\Sph$ the stability exponent (\ref{eq:Sasymptso}) 
of the projection to $\TRd$ entered Theorem~\ref{prop:1stapprox2}. Revisiting
the proof of Proposition~\ref{prop:Lyapfinite} shows that in view of
(\ref{eq:Sspinorbitdef}) such a stability exponent is finite under the
same conditions as stated in Corollary~\ref{cor:Lyapfinite} for 
$\la_{\mathrm{so},k}$. Moreover, a simple estimate yields the bound
\begin{equation*}
 \tilde{\la}_{{\mathrm{max}}} \leq \la_{\mathrm{so},\mathrm{max}} \ . 
\end{equation*}

%% file: main.bbl
\def\cprime{$'$} \def\polhk#1{\setbox0=\hbox{#1}{\ooalign{\hidewidth
  \lower1.5ex\hbox{`}\hidewidth\crcr\unhbox0}}}
\providecommand{\bysame}{\leavevmode\hbox to3em{\hrulefill}\thinspace}
\begin{thebibliography}{SFH{\u{Z}}01}

\bibitem[BB00]{BonDeB00}
F.~Bonechi and S.~De Bi\`evre, \emph{{Exponential mixing and $|$ln $\hbar |$
  time scales in quantized hyperbolic maps on the torus}}, {Commun. Math.
  Phys.} \textbf{211} (2000), 659--686.

\bibitem[BG00]{BolGla00}
J.~Bolte and R.~Glaser, \emph{{Quantum ergodicity for Pauli Hamiltonians with
  spin $1/2$}}, {Nonlinearity} \textbf{13} (2000), 1987--2003.

\bibitem[BG04]{BolGla02}
J.~Bolte and R.~Glaser, \emph{{A semiclassical Egorov theorem and quantum
  ergodicity for matrix valued operators}}, Commun. Math. Phys., to appear,
  arXiv math-ph/0204018, 2004.

\bibitem[BGK01]{BolGlaKep01}
J.~Bolte, R.~Glaser, and S.~Keppeler, \emph{{Quantum and classical ergodicity
  of spinning particles}}, {Ann. Phys. (NY)} \textbf{293} (2001), 1--14.

\bibitem[BK99a]{BolKep99a}
J.~Bolte and S.~Keppeler, \emph{{A semiclassical approach to the Dirac
  equation}}, {Ann. Phys. (NY)} \textbf{274} (1999), 125--162.

\bibitem[BK99b]{BolKep99b}
J.~Bolte and S.~Keppeler, \emph{{Semiclassical form factor for chaotic systems
  with spin $1/2$}}, {J. Phys. A: Math. Gen.} \textbf{32} (1999), 8863--8880.

\bibitem[BM69]{BohMot69}
A.~Bohr and B.~R. Mottelson, \emph{{Nuclear Structure}}, vol.~1, {Benjamin},
  {Reading, Mass.}, 1969.

\bibitem[BR02]{BouRob02}
A.~Bouzouina and D.~Robert, \emph{{Uniform semiclassical estimates for the
  propagation of quantum observables}}, {Duke Math. J.} \textbf{111} (2002),
  223--252.

\bibitem[CFS82]{CorFomSin82}
I.~P. Cornfeld, S.~V. Fomin, and Ya.~G. Sinai, \emph{{Ergodic Theory}},
  {Grundlehren der mathematischen Wissenschaften}, vol. 245, {Springer-Verlag},
  {Berlin, Heidelberg, New York}, 1982.

\bibitem[Chi79]{Chi79}
B.~V. Chirikov, \emph{{A universal instability of many-dimensional oscillator
  systems}}, {Phys. Rep.} \textbf{52} (1979), 264--379.

\bibitem[CR97]{ComRob97}
M.~Combescure and D.~Robert, \emph{{Semiclassical spreading of quantum wave
  packets and applications near unstable fixed points of the classical flow}},
  Asymptot. Anal. \textbf{14} (1997), 377--404.

\bibitem[DS99]{DimSjo99}
M.~Dimassi and J.~Sj\"ostrand, \emph{{Spectral Asymptotics in the
  Semi-Classical Limit}}, {London Mathematical Society Lecture Notes}, vol.
  268, {Cambridge University Press}, {Cambridge}, 1999.

\bibitem[Fol89]{Fol89}
G.~B. Folland, \emph{{Harmonic Analysis in Phase Space}}, {Annals of
  Mathematics Studies}, vol. 122, {Princeton University Press}, Princeton, New
  Jersey, 1989.

\bibitem[Hel75]{Hel75}
E.~J. Heller, \emph{{Time-dependent approach to semiclassical dynamics}}, {J.
  Chem. Phys.} \textbf{62} (1975), 1544--1555.

\bibitem[HJ00]{HagJoy00}
G.~A. Hagedorn and A.~Joye, \emph{{Exponentially accurate semiclassical
  dynamics: propagation, localization, {E}hrenfest times, scattering, and more
  general states}}, {Ann. Henri Poincar{\'e}} \textbf{1} (2000), 837--883.

\bibitem[H{\"o}r90]{Hoe90}
L.~H{\"o}rmander, \emph{{The Analysis of Linear Partial Differential Operators
  I}}, 2nd ed., {Grundlehren der mathematischen Wissenschaften}, vol. 256,
  {Springer-Verlag}, {Berlin, Heidelberg, New York}, 1990.

\bibitem[HPS83]{HogPotSch83}
H.~Hogreve, J.~Potthoff, and R.~Schrader, \emph{{Classical limits for quantum
  particles in external Yang-Mills potentials}}, {Commun. Math. Phys.}
  \textbf{91} (1983), 573--598.

\bibitem[Kep03]{Kep03a}
S.~Keppeler, \emph{{Semiclassical quantisation rules for the {D}irac and
  {P}auli equations}}, {Ann. Phys. (NY)} \textbf{304} (2003), 40--71.

\bibitem[KS85]{KlaSka85}
J.~R. Klauder and B.~S. Skagerstam (eds.), \emph{{Coherent States. Applications
  in Physics and Mathematical Physics}}, {World Scientific}, {Singapore}, 1985.

\bibitem[KW02]{KepWin02}
S.~Keppeler and R.~Winkler, \emph{{Anomalous magneto-oscillations and spin
  precession}}, {Phys. Rev. Lett.} \textbf{88} (2002), 046401.

\bibitem[Laz93]{Laz93}
V.~F. Lazutkin, \emph{{KAM Theory and Semiclassical Approximation to
  Eigenfunctions}}, Ergebnisse der Mathematik und ihrer Grenzgebiete, vol.~24,
  Springer-Verlag, Berlin, Heidelberg, New York, 1993.

\bibitem[Lit86]{Lit86}
R.~G. Littlejohn, \emph{{The semiclassical evolution of wave packets}}, {Phys.
  Rep.} \textbf{138} (1986), 193--291.

\bibitem[Ose68]{Ose68}
V.~I. Oseledec, \emph{{A multiplicative ergodic theorem. Lyapunov
  characteristic numbers for dynamical systems}}, Trans. Moscow Math. Soc.
  \textbf{19} (1968), 197--231.

\bibitem[Per86]{Per86}
A.~Perelomov, \emph{{Generalized Coherent States and Their Applications}},
  Texts and Monographs in Physics, Springer-Verlag, Berlin, Heidelberg, New
  York, 1986.

\bibitem[Rob87]{Rob87}
D.~Robert, \emph{{Autour de l'Approximation Semi-Classique}}, {Progress in
  Mathematics}, vol.~68, {Birkh\"auser}, {Boston, Basel, Stuttgart}, 1987.

\bibitem[SB02]{SilBee02}
P.~G. Silvestrov and C.~P.~J. Beenakker, \emph{{Ehrenfest times for classically
  chaotic systems}}, {Phys. Rev. E.} \textbf{65} (2002), 035208(R).

\bibitem[Sch26]{Sch26}
E.~Schr{\"o}dinger, \emph{{Der stetige {\"U}bergang von der Mikro- zur
  Makromechanik}}, {Naturwissenschaften} \textbf{14} (1926), 664--666.

\bibitem[Sch01]{Sch01}
R.~Schubert, \emph{{Semiclassical localization in phase space}}, Ph.D. thesis,
  {Universit\"at Ulm}, 2001.

\bibitem[Sch04]{Sch04}
R.~Schubert, \emph{{Semiclassical behaviour of expectation values in time
  evolved Lagrangian states for large times}}, {preprint}, 2004, available at
  arXiv:math.MP/0402038.

\bibitem[SFH{\u{Z}}01]{DasFabHuZut01}
S.~Das Sarma, J.~Fabiana, X.~Hua, and I.~{\u{Z}}uti{\'c}, \emph{{Spin
  electronics and spin computation}}, {Solid State Commun.} \textbf{119}
  (2001), 207--215.

\bibitem[Sim80]{Sim80}
B.~Simon, \emph{{The classical limit of quantum partition functions}}, Commun.
  Math. Phys. \textbf{71} (1980), 247--276.

\bibitem[Zas81]{Zas81}
G.~M. Zaslavsky, \emph{{Stochasticity in quantum systems}}, {Phys. Rep.}
  \textbf{80} (1981), 157--250.

\end{thebibliography}
